\documentclass[journal]{IEEEtran}
\usepackage{enumitem}
\usepackage[numbers,sort&compress]{natbib}
\usepackage{amsthm}
\usepackage{amsmath,amssymb,amsfonts}

\usepackage[percent]{overpic}  
\usepackage{colortbl}
\usepackage{array}
\usepackage{algorithmic}
\usepackage{graphicx}
\usepackage{textcomp}
\usepackage{xcolor}

\usepackage{lineno} 
\usepackage{float}

\usepackage{subcaption}

\ifCLASSINFOpdf
 
\else

\fi

\begin{document}

\title{Digital Epidemiology with Awareness-Based Event-Triggered Migration in  Networked Cyber-Physical  Systems}



\author{Yusheng Li,
  Minyu Feng,~\IEEEmembership{Senior Member,~IEEE},
  Liang-jian Deng,~\IEEEmembership{Senior Member,~IEEE},
  Matja\v{z}~Perc,~\IEEEmembership{Member,~IEEE},
  J{\"u}rgen Kurths

  \thanks{

    This work was supported by the Natural Science Foundation of Chongqing under Grant No. CSTB2025YITP-QCRCX0007, by the National Natural Science Foundation of China (NSFC) under Grant No. 62206230, and by the Slovenian Research and Innovation Agency
    under Grant No. P1-0403. \textit{(Corresponding author: Minyu Feng.)}

    Yusheng Li and Minyu Feng are with the College
    of Artificial Intelligence, Southwest University, Chongqing 400715, China
    (e-mail: myfeng@swu.edu.cn).

    Liang-jian Deng is with the School of Mathematical Sciences, University of Electronic Science and Technology of China, Chengdu 611731, China.

    Matja\v{z}~Perc is with the Faculty of Natural Sciences and Mathematics, University of Maribor, 
    Koro{\v s}ka cesta 160, 2000 Maribor, Slovenia,
    with the Community Healthcare Center Dr. Adolf Drolc Maribor, Ulica talcev 9, 2000 Maribor, Slovenia, 
    with the Department of Physics, Kyung Hee University, 26 Kyungheedae-ro, Dongdaemun-gu, Seoul 02447, Republic of Korea, and with the University College, Korea University, 145 Anam-ro, Seongbuk-gu, Seoul 02841, Republic of Korea.

    J{\"u}rgen Kurths is with the Department of Complexity Science, Potsdam Institute for Climate Impact Research, 14473 Potsdam, Germany, and with the Department of Physics, Humboldt University of Berlin, 12489 Berlin, Germany.
  }}


\maketitle

\begin{abstract}

  Understanding how human mobility and  information propagation influence the course of an epidemic remains a key challenge in digital epidemiology. In this work, we develop a new awareness-based, event-triggered epidemic model embedded within a networked Cyber-Physical System (CPS). In our framework, disease transmission and the dissemination of epidemic-related information evolve together on two interconnected layers. 
  In detail, the physical layer models disease spread through human movement between two types of locations -- residences and transfer stations -- forming a bipartite metapopulation network. This structure captures the rendezvous effect, which reflects how gatherings in shared locations contribute to infection spread. The cyber layer represents the flow of information through digital communication networks. We introduce an event-triggered migration regulation mechanism, whereby individuals adapt their movement patterns based on local awareness thresholds, leading to a decentralized control process embedded within the network.
  Using a microscopic Markov chain approach (MMCA), we derive the epidemic threshold analytically and validate our results through extensive Monte Carlo simulations.
  Our findings show that event-triggered migration effectively suppresses the overall spread of the disease and lowers infection peaks -- especially in heterogeneous populations and densely connected gathering points.
  These results demonstrate the potential of CPS-based epidemic models to enable real-time, awareness-driven interventions and to inform the design of decentralized control strategies that leverage digital communication dynamics.

\end{abstract}


\IEEEpeerreviewmaketitle

\section{Introduction}

\IEEEPARstart{T}{he} modeling of infectious disease transmission has undergone decades of development~\cite{pastor2015epidemic}, from the Black Death~\cite{benedictow2004black} to seasonal outbreaks of H1N1~\cite{girard20102009}, drawing sustained interest from scholars seeking to understand contagion processes through mathematical modeling~\cite{ wu2020new}.
With the rise of digital epidemiology~\cite{salathe2012digital}, compartmental models have become a foundational tool in the study of epidemic dynamics~\cite{van2008virus},~\cite{chai2017path}. These models classify individuals into different compartments, such as susceptible (S), exposed (E), infected (I), and recovered (R), and describe the transitions among them using differential equations~\cite{li2021protection},~\cite{xie2023contact}. Classical frameworks like SIS, SIR and SEIR represent the core structure for simulating infection and recovery processes~\cite{tang2025sis}, ~\cite{li2022network}.

Building upon these frameworks, recent studies have focused on incorporating individual social behaviors into epidemic models, particularly behavioral responses influenced by geographic location~\cite{anderson1984spatial}, age~\cite{arenas2020modeling}, social conformity~\cite{pires2017dynamics}, and risk perception~\cite{perra2012activity}.
Techniques such as mean-field approximations~\cite{sahneh2013generalized}, percolation theory~\cite{serrano2006percolation}, and Markov chain approach~\cite{gomez2010discrete} have been used to assess the impact of preventive interventions (e.g., mask-wearing, social distancing, or vaccination) on the evolution of outbreaks~\cite{Mei2026}.
Importantly, these behaviors are shaped by both real-time infection dynamics and rising awareness in local communities, emphasizing the role of self-protective behavior in disease containment~\cite{XIE2024115289},~\cite{esquivel2018efficiency}.

To better explore the spatiotemporal characteristics of epidemic contagion, Granell et al.~\cite{granell2013dynamical} introduced a pioneering multiplex network framework that integrated epidemic spreading and information diffusion across layered structures.
Benefiting from the advances in digital communication, online platforms (e.g., Instagram, Weibo, and TikTok) have largely supplanted physical contact as the dominant channel for epidemic-related information~\cite{feng2023impact},~\cite{zhu2023epidemic}. This development has created dynamic feedback loops among awareness formation, protective action, and disease progression~\cite{yuan2025impacts},~\cite{Pang2025}.
The coupling between information diffusion and epidemic dynamics, which is often termed information–epidemic co-evolution~\cite{xia2019new},~\cite{kolok2025epidemic}, has led to the rise of CPSs, where cyber communication layers interact with physical transmission networks~\cite{chen2021medical}.
In such systems, a one-to-one correspondence exists between nodes in different layers, yet their structural and functional roles remain distinct.
Empirical and theoretical studies have shown that information diffusion can significantly raise epidemic thresholds~\cite{li2021epidemic}, reduce outbreak peaks~\cite{djilali2020coronavirus}, and change the spatial–temporal trajectory of outbreaks~\cite{lion2016spatial}.

Parallel to these advances, the role of human mobility in shaping epidemic dynamics has received growing attention~\cite{bajardi2011dynamical}. Contemporary disease spreading is inherently spatiotemporal, driven by complex patterns of movement among structured populations~\cite{colizza2007reaction},~\cite{soriano2022modeling}.
The metapopulation framework represents regions or communities as interconnected nodes (or patches/subpopulations), linked by recurrent mobility patterns such as commuting and travel~\cite{wang2020network}.
This mobility network topology greatly alters disease propagation mechanisms: while local transmission follows reaction dynamics through direct contacts within patches, global spread emerges from diffusion processes along transportation pathways
\cite{gomez2018critical}.
By incorporating heterogeneous mobility rates and movement pathways, metapopulation models have become vital tools for evaluating travel restrictions, targeted quarantines, and other control measures~\cite{an2024coupled}.

Despite notable advancements, a critical limitation remains: most existing frameworks assume fixed or periodic mobility patterns, neglecting behavioral adaptations that arise in response to perceived risk. In real-world outbreaks, individuals often adjust their mobility decisions dynamically to avoid high-risk areas or alter routines based on rising awareness or digital alerts.
Furthermore, prior studies frequently overlook the impact of structured aggregation behaviors, such as daily commuting between residential zones and shared public hubs. These hubs, including offices, markets, or stations, function as high-density contact points and naturally give rise to the so-called \textit{rendezvous effect}~\cite{cao2011rendezvous}, significantly amplifying inter-patch transmission potential.
To address this gap, we propose a novel awareness-based event-triggered
epidemic model situated in a metapopulation CPS.
Our framework introduces a migration mechanism that activates when awareness surpasses a critical threshold, prompting individuals to break from routine movement and disperse adaptively.
We formalize the system as a bipartite cyber-physical network, consisting of two types of locations, residences and transfer stations, and derive the epidemic threshold using the MMCA. Simulation results validate the theoretical predictions and demonstrate how awareness-driven responses reshape both the intensity and spatial footprint of epidemic outbreaks.

To summarize, the principal contributions of this work are as follows.
\begin{enumerate}[label=\arabic*),topsep=0pt]
  \item We propose a cyber–physical epidemic framework with awareness-based event-triggered migration. The model couples information diffusion in a cyber layer and disease transmission in a physical layer, structured as a bipartite metapopulation network of residences and transfer stations. Unlike prior models focusing on individual-level dynamics, our patch-level awareness formulation reduces computational complexity while preserving analytical tractability.

  \item We introduce a novel migration mechanism triggered by local awareness levels. When the proportion of aware individuals in a patch exceeds a defined threshold, individuals adaptively deviate from the routine mobility, reflecting realistic behavioral responses to outbreaks.

  \item We derive the epidemic outbreak threshold analytically for the coupled cyber–physical system using the MMCA. Our framework captures the emergent rendezvous effect, where individuals from distinct origins converge at shared hubs, offering new insights into how awareness diffusion reshapes both epidemic peaks and spatial spread. The theoretical predictions are validated by extensive MC simulations and further benchmarked against classical baseline models, highlighting the distinct advantage of awareness-triggered migration in suppressing epidemic prevalence.

\end{enumerate}

The rest of the paper is organized as follows.
In Section~\ref{sec2:model description}, we describe the construction process of our epidemic model and its compartmental structure.
In Section~\ref{sec3:threshold}, we derive the epidemic threshold analytically at two different types of locations using the MMCA.
In Section~\ref{sec4:numerical_simulation}, we validate the model with extensive MC simulations and analyze the impact of awareness-driven migration and coupled spreading processes on the epidemic dynamics.
Finally, we conclude the findings and outline directions for future research in Section~\ref{sec5:conclusion}.

\section{Epidemic Model Coupling Awareness in Multiplex Metapopulation Networks}
\label{sec2:model description}
This section outlines the formulation of our epidemic model, which captures the coupled dynamics of disease transmission and information diffusion on a structured CPS. We describe the model architecture, compartmental transitions, and the mechanism of event-triggered migration. Table ~\ref{tab:notation} summarizes the model parameters and variables throughout this
article.

\begin{table*}[!t]
  \centering
  \caption{Summary of key variables and parameters in the coupled awareness--epidemic metapopulation model.}
  \label{tab:notation}
  \renewcommand{\arraystretch}{1.2}
  \definecolor{lightgray}{gray}{0.9}
  \rowcolors{2}{white}{lightgray}

  \begin{tabular}{>{\centering}m{2.8cm} >{\centering}m{3.2cm} >{\raggedright\arraybackslash}m{8.5cm}}

    \hline
    \multicolumn{1}{c}{\textbf{Symbol}}              &
    \multicolumn{1}{c}{\textbf{Layer / Category}}    &
    \multicolumn{1}{c}{\textbf{Definitions}}                                                                                                                                                                                 \\
    \hline

    $N, M$                                           & Network indices        & Number of subpopulations (locations) in the metapopulation network (\(N\) residences
    and \(M\) transfer stations).                                                                                                                                                                                                     \\

    $n_i(t), m_j(t)$                                 & Network indices        & Current number of individuals in residence $i$ and transfer station $j$ at time $t$.                                                              \\

    $i, j, k$                                        & Network indices        & Indices of residences ($i,k=1,\dots,N$) and transfer stations ($j=1,\dots,M$).                                                                    \\

    $W^v = \{w_{ij}^v\}$                             & Cyber virtual layer    & Weight matrix of virtual communication, representing information exchange intensity between patches.                                              \\

    $W^c = \{w_{ij}^c\}$                             & Physical contact layer & Bipartite connectivity matrix between residences and transfer stations, encoding commuting and contact strength.                                  \\

    $R_{ij}$                                         & Migration preference   & Probability of individuals from residence $i$ moving to transfer station $j$.                                                                     \\

    $U, A$                                           & Awareness states       & Unaware and aware states in the cyber layer following the UAU process.                                                                            \\

    $S, I, R$                                        & Epidemic states        & Susceptible, infected, and recovered states in the physical layer following the SIR process.                                                      \\

    US, AS, AI, UR, AR                               & Composite states       & Composite awareness--epidemic states describing individual-level coupling between the two layers.                                                 \\

    $p_i^{X}(t)$                                     & State probability      & Probability that an individual in residence $i$ is in composite state $X$ at time $t$.                                                            \\

    $\lambda, \mu_1$                                 & Information parameters & Information transmission rate and information forgetting rate.                                                                                    \\

    $\beta, \mu_2, \sigma$                           & Epidemic parameters    & Disease transmission rate, recovery rate of infected individuals, and $\sigma$ quantifies awareness-induced protection ($\beta^A=\sigma\beta^U$). \\

    $\theta_i, g, g_i $                             & Mobility parameters    & Location-specific heterogeneity, global mobility baseline and residence-specific mobility propensity (\(g_i = \theta_i g\)).                      \\

    $n_{i\rightarrow i}(t), n_{k\rightarrow j}(t) $ & Mobility parameters    & Number of individuals staying at residence $i$ and number of individuals from residence $k$ moving to transfer station $j$ at time $t$.           \\

    $Q_i^U, Q_i^A$                                   & Infection risk         & Infection probabilities for unaware and aware susceptible individuals  associated with residence $i$ who are susceptible get infected             \\

    $N_i^Y, M_j^Y$                                   & Local / Hub infection  & Infection probabilities in residences ($N_i^Y$) and transfer stations ($M_j^Y$), with $Y\in\{U,A\}$.                                              \\

    $\alpha$                                         & Awareness threshold    & Activation threshold triggering awareness-based adaptive migration.                                                                               \\

    $\varepsilon_i(t)$                               & Migration trigger      & Event-triggered migration intensity controlled by local awareness level.                                                                          \\

    $T_{ij}$                                         & Redistribution matrix  & Probability governing non-local return and redistribution after interaction at transfer stations.                                                 \\

    \hline
  \end{tabular}
\end{table*}

\subsection{Architecture of the Networked Cyber-Physical System}
We consider a two-layer metapopulation networked CPS that represents structured populations.
The cyber virtual layer models the diffusion of information across subpopulations via digital communication platforms such as WeChat or TikTok.
The physical contact layer is designed as a bipartite network connecting two types of nodes, residences (e.g., homes or dormitories) and transfer stations (e.g., workplaces, commercial centers). This structure emulates the rendezvous effect, where individuals from different residences meet at transfer stations, creating transient yet dense mixing environments. These rendezvous nodes are pivotal in facilitating both epidemic and information diffusion, particularly in urban settings with structured commuting behavior.
Each node in the network corresponds to a patch, and edges are the connections among patches.

\begin{figure*}[!t]
  \centering
  \begin{overpic}[width=0.95\linewidth]{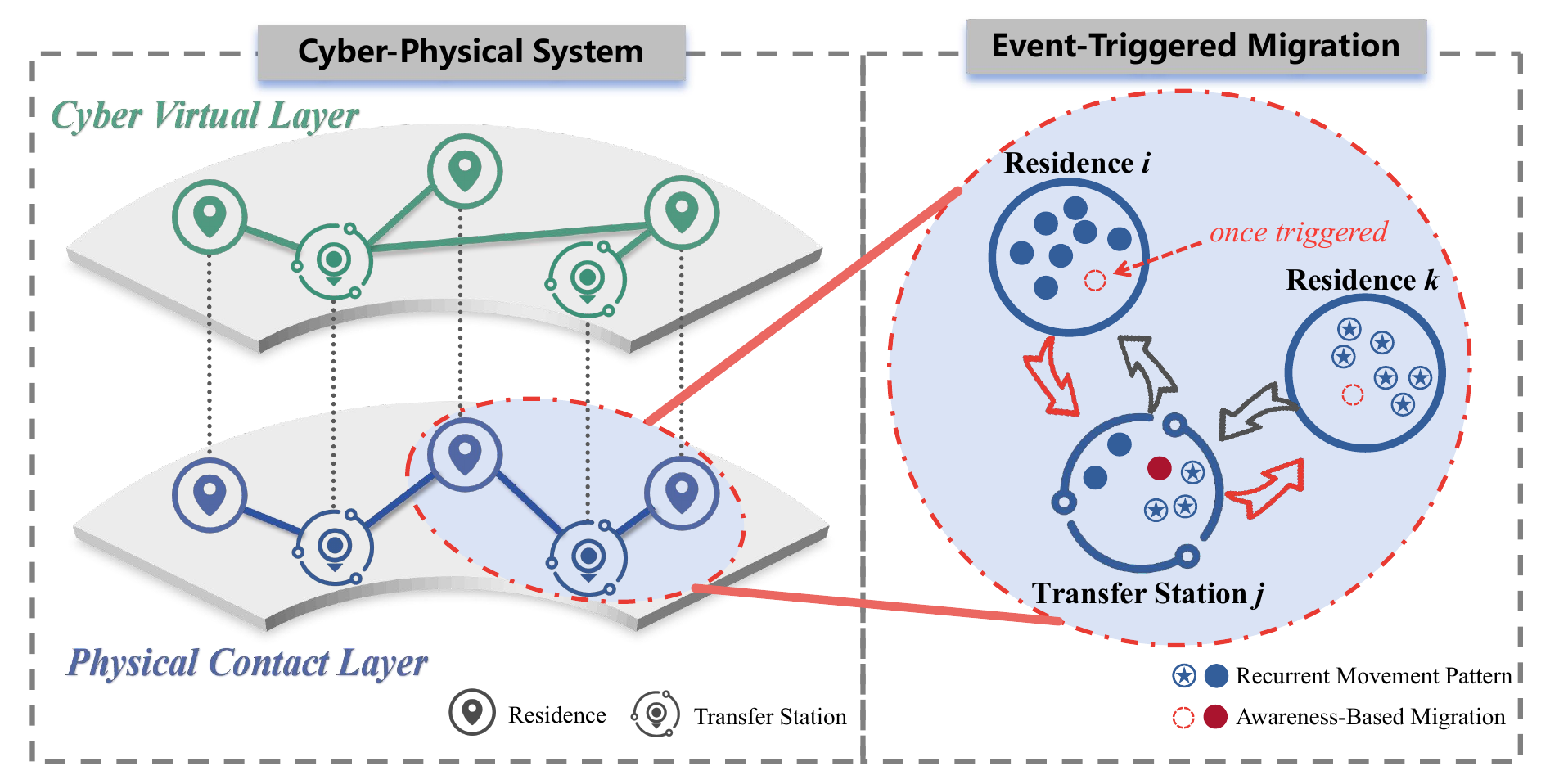}
    \put(28,-0.7){\textbf{(a)}}
    \put(75,-0.7){\textbf{(b)}}
  \end{overpic}
  \caption{ \textbf{Architecture of the metapopulation networked CPS and schematic of event-triggered migration mechanism. }
    (a) Two-layered cyber-physical metapopulation system. Upper cyber layer for information dissemination and lower physical layer for epidemic transmission, connected via bipartite topology with residences (circles with address markers) and transfer stations (circles with outer rings).
    (b) Event-triggered migration. When local awareness level in residence $i$ exceeds a threshold, individuals (red) deviate from recurrent movement patterns (blue) and migrate from residence $i$ to $k$ via transfer station $j$.
  }
  \label{fig:multiplex_network}
\end{figure*}

Fig.~\ref{fig:multiplex_network} presents the conceptual framework of our metapopulation cyber-physical system. The architecture of Fig.~\ref{fig:multiplex_network}(a) integrates two interconnected layers: a cyber layer for information flow regarding epidemic risks and a physical layer for disease transmission through human contacts. This design captures the co-evolution of information awareness and epidemic dynamics. Individuals are associated with specific residential patches and can access both layers, enabling bidirectional influence between awareness states and infection states.
The bipartite topology organizes nodes into residences and transfer stations, structuring human mobility within and between patches. A key feature is the event-triggered migration mechanism shown in Fig.~\ref{fig:multiplex_network}(b), where individuals dynamically adjust their movement based on local awareness conditions. When the awareness level within a residence crosses an activation threshold, residents may initiate protective migration to other residences via transfer stations, reflecting adaptive behavioral responses to perceived epidemic threats. This mechanism creates a dynamic coupling between information dissemination in the cyber layer and mobility behaviors in the physical layer.


For the information diffusion in the cyber layer, the traditional UAU awareness model is adopted to capture the dissemination of disease information.
Each individual associated to its patch is either unaware of the disease (U-state) or aware about the epidemic spreading (A-state).
Unaware individuals may transition to the A-state under two conditions: (1) upon infection or (2) after receiving disease-related information through social interactions (e.g., conversations with friends or online activity). Conversely, aware individuals may revert to the U-state after recovery or over time due to waning vigilance.
The unaware–aware–unaware (UAU) process is characterized through the following equations:
\begin{equation}
  U + A \xrightarrow{\lambda} 2 A, \quad A \xrightarrow{\mu_1} U,
\end{equation}
where the equation identifies that an unaware (U) individual interacts with an aware (A) individual and becomes aware at the information transmission rate \( \lambda \), and an aware (A) individual loses awareness and returns to the unaware (U) state at information forgetting rate \( \mu_1 \).

We assume that information is uniformly distributed within each patch and that the cyber virtual network, which represents interactions between individuals and their neighbors remains static throughout the evolution of the system. This simplification is justified by the relative stability of online social connections compared to the dynamic physical contacts driven by human mobility. Long-term social ties on digital platforms exhibit minimal variation, whereas spatial proximity in the physical layer changes frequently.
The inter-patch communication intensity is encoded by the virtual-layer weight matrix \(W^v={(w_{ij}^v)}\), where \(w_{ij}^v\)
quantifies the frequency of information exchange between patches \(i\) and \(j\). Local interactions are captured by diagonal elements \(w_{ii}^v\), while off-diagonal terms \(w_{ij}^v(i\ne j)\) represent non-local digital communication.

In the physical layer, to better mimic the rendezvous effects of real world scenarios, such as meetings and social gatherings, we introduce a bipartite network considering two kinds of locations: $N$ residences and $M$ transfer stations.
For residences, such as apartments and houses, there are $n_i$ individuals, where $i=1,2,\cdots, N$.
Analogous to the awareness model in the cyber layer, we adopt the SIR framework to describe the transmission dynamics between susceptible and infected individuals. The transitions can be formalized as follows:
\begin{equation}
  S + I \xrightarrow{\beta} 2 I, \quad I \xrightarrow{\mu_2} R.
\end{equation}
When a susceptible individual (S) interacts with an infected individual (I), the susceptible individual becomes infected at disease spreading rate \( \beta \), which results in two infected individuals (2I). Infected individuals (I) recover with recovery rate \( \mu_2 \), transitioning to the recovered state (R).

The structural coupling between residences and transfer stations is defined by a physical-layer weight matrix \(W^c = (w_{ij}^c)\), where \(w_{ij}^c\) captures the interaction strength (e.g., commuting frequency, travel accessibility) between residence \(i\) and transfer station \(j\). No intra-layer edges exist within the same group (i.e., between residences or between stations), ensuring a pure bipartite structure.
These weights also govern both individual movement and contact frequency. Specifically, we define the probability of an individual from residence \(i\) moving to transfer station \(j\) as:
\begin{equation}
  R_{ij} = \frac{w_{ij}^c}{\sum_{m=1}^M w_{im}^c}.
  \label{eq:Rij}
\end{equation}
Here, \(R_{ij}\) represents a normalized migration preference that reflects both physical connectivity and travel intensity. This formulation supports analytical tractability and forms the basis for modeling awareness-based migration in later sections.

It is important to note that the transfer stations act as hubs for frequent interactions and do not have a fixed population. These interactions are represented using a well-mixed approximation, a common approach in existing studies, to enable rigorous theoretical analysis.
The movement in the bipartite network follows a strictly sequential process, in which individuals must first transition from a residence to a transfer station before proceeding to another residence or returning to their origin.

\subsection{Coupled Compartmental Dynamics}

To investigate the co-evolution dynamics of epidemic propagation and information diffusion in the metapopulation networked CPS, we construct a coupled UAU-SIR model using the MMCA.

During the spreading process of the epidemic, each individual in the networked CPS is classified into one of five composite states: US (unaware and susceptible), AS (aware and susceptible), AI (aware and infected), UR (unaware and recovered), and AR (aware and recovered). We denote the probability of an individual in patch $i$ being one of these states at time $t$ as $p_i^{US}(t)$, $p_i^{AS}(t)$, $p_i^{AI}(t)$, $p_i^{UR}(t)$ and $p_i^{AR}(t)$, respectively. Here, a timestep \(t\) represents a unit of time, simulating a day in daily life.
Once an unaware healthy individual (US) gets infected, it will be conscious of the disease immediately and transit to the AI state, which means the UI state does not apparently exist. Thus, we can get the normalization equation as
\begin{equation}
  p_i^{US}(t) +  p_i^{AI}(t) + p_i^{AS}(t)+  p_i^{UR}(t) + p_i^{AR}(t) \equiv  1 .
  \label{eq:norm}
\end{equation}

Single-layer state probabilities are expressed as
\begin{equation}
  \left\{
  \begin{array}{l}
    p_{i}^{A}(t)=p_{i}^{A S}(t)+p_{i}^{A I}(t)+p_{i}^{A R}(t) \\
    p_{i}^{U}(t)=p_{i}^{U S}(t)+p_{i}^{U R}(t)                \\
    p_{i}^{I}(t)=p_{i}^{A I}(t)                               \\
    p_{i}^{S}(t)=p_{i}^{A S}(t)+p_{i}^{U S}(t)                \\
    p_{i}^{R}(t)=p_{i}^{A R}(t)+p_{i}^{U R}(t).
  \end{array}
  \right.
  \label{eq:single_states}
\end{equation}

Although transfer stations lack permanent populations, they host transient individuals from multiple residences. During occupancy, these individuals exchange epidemic-related information, prompting us to define an effective transient awareness ratio in transfer station $j$ as
\begin{equation}
  \tilde{p}_{j}^{A}(t)=\frac{\sum_{k=1}^{N} p_{k}^{A}(t) n_{k \rightarrow j} (t)}{\sum_{k=1}^{N} n_{k \rightarrow j}(t)},
\end{equation}
where \(n_{k \rightarrow j}\) is the number of individuals from residence $k$ moving to transfer station $j$, which can be expressed as
\begin{equation}
  n_{k\rightarrow j}(t)=n_k(t) \  g_k \ R_{kj}.
  \label{eq:n_k-j}
\end{equation}

In the cyber layer, the probability that individuals in residence \(i\) are informed due to their neighbors is given by
\begin{equation}
  r_i(t)=1-\prod_{j=1}^{M}  \left [ 1-\lambda w_{ji}^v\tilde{p}_j^A(t) \right ].
\end{equation}

Furthermore, the probability $Q_i^U(t)$ ($Q_i^A(t)$) indicates unaware (aware) individuals associated with  residence $i$ who are susceptible  get infected at the end of time $t$, which is formalized as
\begin{equation}
  Q_i^U(t)=(1-g_i)N_i^U(t)+g_i\sum_{j=1}^{M}R_{ij} M_j^U(t)
  \label{eq:Q_i^U}
\end{equation}
and,
\begin{equation}
  Q_i^A(t)=(1-g_i)N_i^A(t)+g_i\sum_{j=1}^{M}R_{ij} M_j^A(t).
  \label{eq:Q_i^A}
\end{equation}
The two terms in Eq.~(\ref{eq:Q_i^U}) include the fraction of unaware individuals that do not move and expose to the disease-carriers getting infected in patch $i$, and the fraction of the unaware population traveling to the connected transfer stations and getting infected. Eq.~(\ref{eq:Q_i^A}) follows the same structure but applies to aware individuals.

The formulations for the probabilities of being infected to unaware and aware individuals in (but not necessarily
associated with) residence $i$ are, respectively,
\begin{equation}
  N_i^U(t)=1-\left [ 1-\beta^U p_i^I(t) \right ] ^{n_{i\rightarrow i} (t)}
  \label{eq:N_i^U}
\end{equation}
and,
\begin{equation}
  N_i^A(t)=1-\left [ 1-\beta^A p_i^I(t) \right ] ^{n_{i\rightarrow i} (t)}
  \label{eq:N_i^A}
\end{equation}
where unaware susceptible individuals (US) are infected at a rate \( \beta^U \) and aware susceptible individuals (AS) are infected at a rate \( \beta^A \). We assume that \(\beta^U = \beta\), and \(\beta^A = \sigma \beta\) with \(0 \le \sigma \le 1\) denoting awareness-induced protection. The number of individuals staying at residence $i$ is
\begin{equation}
  n_{i\rightarrow i}(t)=n_i(t)(1-g_i).
  \label{eq:n_i-i}
\end{equation}

The probabilities of unaware or aware individuals in transfer station $j$ being infected are calculated as
\begin{equation}
  M_j^U(t)=1-\prod_{k=1}^{N} \left [ 1-\beta^U p_k^I(t) \right ] ^{n_{k\rightarrow j}(t) }
  \label{eq:M_j^U}
\end{equation}
and,
\begin{equation}
  M_j^A(t)=1-\prod_{k=1}^{N} \left [ 1-\beta^A p_k^I(t) \right ] ^{n_{k\rightarrow j}(t) }
  \label{eq:M_j^A}
\end{equation}
where \(n_{k\rightarrow j}\) is the floating population explained in Eq. (\ref{eq:n_k-j}).

\begin{figure}[htbp]
  \centering
  \includegraphics[width=0.48\textwidth]{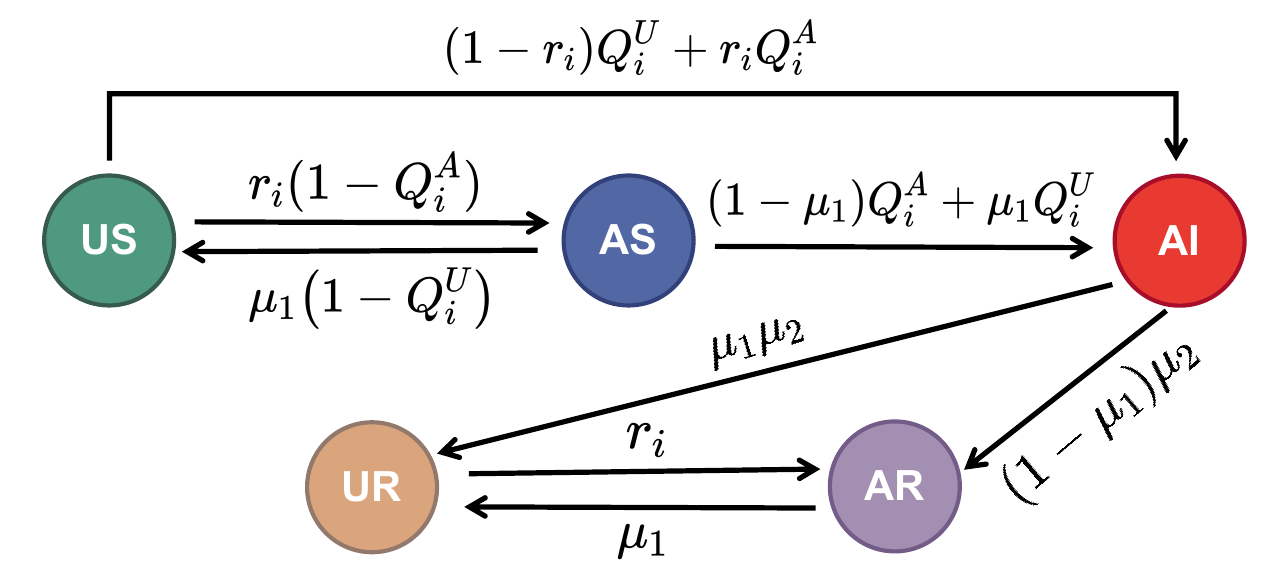}
  \caption{\textbf{State transitions among five composite compartments in the UAU–SIR model on a metapopulation CPS.}
    The diagram illustrates the probabilistic transitions of individuals located in residence $i$ across five composite compartments within a single time step. Each state is defined by a pair of epidemic and awareness states: US (unaware and susceptible), AS (aware and susceptible), AI (aware and infected), UR (unaware and recovered), and AR (aware and recovered). Transition mechanisms are governed by multiple parameters: $\mu_1$ is the information forgetting rate, $\mu_2$ is the disease recovery rate, $r_i$ denotes the probability that an unaware individual in residence $i$ is informed by neighbors, $Q_i^U$ is the infection probability for unaware susceptible individuals in (but not necessarily associated with) $i$, and $Q_i^A$ is the infection probability for aware susceptible individuals.}

  \label{fig:state}
\end{figure}

Combining the elaboration above, the state transition process of the five compartmental states is shown in Fig.~\ref{fig:state}, and we obtain the co-evolution dynamics of our coupled metapopulation model according to the MMCA as
\begin{equation}
  \left\{
  \begin{aligned}
    p_{i}^{U S}(t+1)  & = p_{i}^{U S}(t)\left[1-r_{i}(t)\right]\left[1-Q_{i}^{U}(t)\right]                       \\&
    +p_{i}^{A S}(t)\mu_1\left[1-Q_{i}^{U}(t)\right]                                                              \\
    p_{i}^{A S}(t+1)  & = p_{i}^{A S}(t)\left(1-\mu_{1}\right)\left[1-Q_{i}^{A}(t)\right]                        \\&
    +p_{i}^{U S}(t) r_{i}(t)\left[1-Q_{i}^{A}(t)\right]                                                          \\
    p_{i}^{A I}(t+1)  & = p_{i}^{A I}(t)\left(1-\mu_{2}\right)+ p_{i}^{A S}(t)\left(1-\mu_{1}\right)Q_{i}^{A}(t) \\&
    +p_{i}^{A S}(t) \mu_{1} Q_{i}^{U}(t) +p_{i}^{U S}(t)\left[1-r_{i}(t)\right] Q_{i}^{U}(t)                     \\&
    +p_{i}^{U S}(t) r_{i}(t)Q_{i}^{A}(t)
    \\
    p_{i}^{ U R}(t+1) & = p_{i}^{A I}(t) \mu_{1} \mu_{2}+p_{i}^{A R}(t) \mu_{1}                                  \\&
    +p_{i}^{U R}(t)\left[1-r_{i}(t)\right]
    \\
    p_{i}^{A R}(t+1)  & = p_{i}^{A R}(t)\left(1-\mu_{1}\right)+p_{i}^{A I}(t)\left(1-\mu_{1}\right) \mu_{2}      \\&
    +p_{i}^{U R}(t)r_{i}(t) .
  \end{aligned}
  \right.
  \label{eq:MMCA}
\end{equation}

\subsection{Event-Triggered Migration Mechanism}

To capture the dynamic influence of awareness on population mobility during epidemic outbreaks, we introduce an event-triggered migration mechanism as an extension to the classic Movement–Interaction–Return (MIR) framework~\cite{soriano2018spreading}. This mechanism operates through two coupled dynamic processes: First, individual awareness levels evolve continuously via the UAU process during recurrent movement, changing the aware fraction within each patch over time. Second, and crucially, once the aggregate awareness in a residence surpasses a predefined threshold, it triggers a behavioral shift. This initiates our proposed awareness-based migration, where individuals deviate from their regular return pattern and are redistributed to other residences. This redistribution dynamically alters patch population compositions, thereby creating a feedback loop between mobility and the cyber layer's awareness states.

This mechanism unfolds over four sequential stages:
(1) awareness activation,
(2) heterogeneous mobility initiation,
(3) interaction at rendezvous nodes, and
(4) non-local redistribution and iteration.
The following subsections detail each phase of the process.

\subsubsection{Awareness Activation}
At the beginning of an outbreak, individuals in each residence are either unaware (U) or aware (A) of the disease, based on the UAU model.
Once the local proportion of aware individuals in residence \(i\) exceeds an activation threshold \(\alpha\), the migration process is triggered. This  activation factor is modeled as
\begin{equation}
  \varepsilon_i(t) = H\left(p_i^A(t) - \alpha\right) \varepsilon_0.
\end{equation}
\(H(x)\) denotes the Heaviside step function where  \( H(x) =1 \)  if \( x>0 \) , \( H(x) =0 \) otherwise. \(\varepsilon_0\) represents the activation intensity of awareness-based behavioral response.

\subsubsection{Heterogeneous Mobility Initiation}

When migration is triggered by local awareness, the total number of individuals flowing out from residence \(i\) at time \(t\) is given by:
\begin{equation}
  F_i(t) = n_i(t) g_i \varepsilon_i(t) ,
\end{equation}
where individuals from residence \(i\) initiate movement toward surrounding transfer stations with a probability defined as \(g_i = \theta_i g\), where \(g\) denotes the global mobility baseline and \(\theta_i \in [0,1]\) reflects the location-specific heterogeneity, e.g., socioeconomic factors, geographic accessibility, or local policy constraints.
\(n_i(t)\) represents the current population size of residence \(i\).

This formulation ensures that only patches experiencing high awareness levels initiate adaptive mobility, distinguishing from
recurrent movement patterns.

\subsubsection{Interaction at Rendezvous Nodes}
After migration, individuals gather at transfer stations—interpreted as rendezvous nodes such as workplaces, malls, or transit hubs—where physical contact and disease transmission occur. The total population present at transfer station \(j\) is calculated as:
\begin{equation}
  m_j(t) = \sum_{i=1}^{N} n_i(t) g_i \varepsilon_i(t) R_{ij}.
\end{equation}
If awareness does not reach the activation threshold in any residence, no migration takes place and the model reverts to the conventional MIR structure with zero net movement at the end of the time step.

\subsubsection{Non-Local Return and Iteration}
Following the interaction phase, individuals do not necessarily return to their original residences. Instead, their return is governed by a redistribution matrix:
\begin{equation}
  T_{ij} = \frac{w_{ij}^c}{\sum_{n=1}^N w_{in}^c}.
\end{equation}
The total number of individuals returning to residence \(i\) is:
\begin{equation}
  R_i(t) = \sum_{j=1}^{M} m_j(t) T_{ij}.
\end{equation}
This return mechanism reflects behavioral reallocation rather than fixed recurrent movement, allowing individuals to redistribute based on contact intensity and spatial structure. The updated population size of residence \(i\) at time \(t+1\) is thus:
\begin{equation}
  \begin{aligned}
    n_i(t+1) & = n_i(t) - F_i(t) + R_i(t)                                             \\
             & = n_i(t) - n_i(t) g_i \varepsilon_i(t) + \sum_{j=1}^{M} m_j(t) T_{ij}.
  \end{aligned}
\end{equation}

This four-stage mechanism captures how local awareness dynamically shapes spatial movement and interaction, reinforcing the role of cyber-layer information in mitigating the physical-layer contagion process.

\section{Theoretical Analysis of Epidemic Threshold}
\label{sec3:threshold}

In this section, we analyze the epidemic threshold of the proposed metapopulation networked CPS by utilizing MMCA dynamic equations.

When the entire system of coupled dynamics reaches a steady state (\(t\rightarrow \infty \)), the evolution of each state $X$ mentioned in Eq. (\ref{eq:MMCA}) is given by
\begin{equation}
  p_{i}^{X}=\lim _{t \rightarrow \infty} p_{i}^{X}(t),
\end{equation}
where the state \(X\) represents the state of \(US\), \(AI\), \(UR\), \(AS\) or \(AR\), respectively.

With the assumption that near the epidemic threshold, the epidemic process is in its incipient stage,
where infection either dies out or grows from an infinitesimal seed.
Therefore, the proportion of infected individuals remains vanishingly small compared to the total population.
In this regime, the MMCA equations can be linearized around the disease-free equilibrium,
which is a standard approach for deriving epidemic thresholds in metapopulation and multiplex spreading models.
Hence, we have
$p_{i}^{AI}(t), p_{i}^{UR}(t), p_{i}^{AR}(t) \ll 1$
and
$Q_{i}^{U}(t), Q_{i}^{A}(t), N_{i}^{U}(t),N_{i}^{A}(t) \rightarrow 0$. Consequently, we obtain the abbreviated formulations of Eq. (\ref{eq:MMCA}) in the steady state as
\begin{equation}
  \left\{
  \begin{aligned}
    p_{i}^{U S} & =p_{i}^{U S}\left(1-r_{i}\right)+p_{i}^{A S} \mu_{1}                                            \\
    p_{i}^{A S} & =p_{i}^{A S}\left(1-\mu_{1}\right)+p_{i}^{U S} r_{i}                                            \\
    p_{i}^{A I} & =p_{i}^{A I}\left(1-\mu_{2}\right)+p_{i}^{A S}\left(1-\mu_{1}\right)Q_{i}^{A}                   \\&
    +p_{i}^{A S} \mu_{1}  Q_{i}^{U}+p_{i}^{U S}\left(1-r_{i}\right)Q_{i}^{U}+p_{i}^{U S} r_{i}Q_{i}^{A}           \\
    p_{i}^{U R} & =p_{i}^{A I} \mu_{1} \mu_{2}+p_{i}^{A R} \mu_{1}+p_{i}^{U R}\left(1-r_{i}\right)                \\
    p_{i}^{A R} & =p_{i}^{A R}\left(1-\mu_{1}\right)+p_{i}^{A I}\left(1-\mu_{1}\right) \mu_{2}+p_{i}^{U R} r_{i}.
  \end{aligned}
  \right.
  \label{eq:MMCA-new}
\end{equation}

Denote the small infected probability close to the critical point as
\begin{equation}
  p_i^{AI} = \epsilon_i^* \ll 1.
\end{equation}
Since infected individuals can only arise from susceptible compartments through the infection terms
$Q_i^U$ and $Q_i^A$, the steady-state balance of the $AI$ equation in Eq.~(\ref{eq:MMCA-new})
yields that the recovery flow $\mu_2 \epsilon_i^*$ must equal the total newly generated infections.
By collecting all first-order contributions in $\epsilon_i^*$, we obtain
\begin{equation}
  \begin{aligned}
    \mu_2 \epsilon_i^* & = Q_i^A \left[ p_i^{AS}(1 - \mu_1) + p_i^{US}r_i \right] \\&
    + Q_i^U \left[ p_i^{AS} \mu_1 + p_i^{US}(1 - r_i) \right]                     \\
                       & = Q_i^A p_i^{AS} + Q_i^U p_i^{US}.
  \end{aligned}
  \label{eq:mu2-epsilon_i}
\end{equation}

Neglecting higher--order terms of $\epsilon_i^{*}$ corresponds to retaining only linear contributions
in the early outbreak regime.
In particular, because $\epsilon_i^{*}$ is infinitesimal, the binomial approximation
$(1-\epsilon)^n \approx 1-n\epsilon$ holds for large but finite patch populations,
allowing us to express the infection probabilities $N_i^Y$ and $M_j^Y$
as linear functions of $\epsilon_i^{*}$ where the sign \(Y\) represents the U-state or A-state as
\begin{equation}
  \begin{aligned}
    N_{i}^{U} \approx & n_{i\rightarrow  i} \beta^{U} p_{i}^{I}=n_{i}(1-g_i)  \beta^{U} \epsilon_{i}^{*}                                      \\
    M_{j}^{U} \approx & \sum_{k=1}^{N} n_{k \rightarrow j} \beta^{U} p_{k}^{I} =\sum_{k=1}^{N} n_{k} g_k  R_{k j}  \beta^{U} \epsilon_{k}^{*}
  \end{aligned}
  \label{eq:NM_approximation1}
\end{equation}
and,
\begin{equation}
  \begin{aligned}
    N_{i}^{A} \approx & n_{i \rightarrow i} \beta^{A} p_{i}^{I}=n_{i}(1-g_i) \beta^{A} \epsilon_{i}^{*}                                       \\
    M_{j}^{A} \approx & \sum_{k=1}^{N} n_{k\rightarrow j} \beta^{A} p_{k}^{I}=\sum_{k=1}^{N} n_{k} g_k R_{k j}  \beta^{A} \epsilon _{k}^{*} .
  \end{aligned}
  \label{eq:NM_approximation2}
\end{equation}

Substituting the Eqs. (\ref{eq:n_k-j}), (\ref{eq:n_i-i}), (\ref{eq:NM_approximation1}) and (\ref{eq:NM_approximation2}) into the formulations of \(Q_i^A\) and \(Q_i^U\), we get
\begin{equation}
  \begin{aligned}
    Q_i^U & = \beta^U \sum_{k=1}^N \left[ (1 - \theta_i g)^2 n_i \delta_{ik} + \theta_i \theta_k g^2 \sum_{j=1}^M R_{ij} R_{kj} n_k\right] \epsilon_k^*     \\
    Q_i^A & = \beta^A \sum_{k=1}^N \left[ (1 - \theta_i g)^2 n_i \delta_{ik} + \theta_i \theta_k g^2 \sum_{j=1}^M R_{ij} R_{kj} n_k \right] \epsilon_k^*  .
  \end{aligned}
\end{equation}

For the sake of simplicity, we define
\begin{equation}
  M_{ik} = (1 - \theta_i g)^2 n_i \delta_{ik} + \theta_i \theta_k g^2 \sum_{j=1}^M R_{ij} R_{kj} n_k,
  \label{eq:M_ik}
\end{equation}
where the first term $(1-\theta_i g)^2 n_i \delta_{ik}$ captures purely local infections
generated by individuals remaining in residence $i$,
while the second term explicitly accounts for cross-residence transmission pathways
mediated by shared transfer stations.
Therefore, $M_{ik}$ can be interpreted as an effective mobility-induced mixing kernel,
quantifying how infections originating from residence $k$ contribute to the risk in residence $i$.

Then, we further obtain
\begin{equation}
  \begin{aligned}
    Q_{i}^{U} & =\beta^{U} \sum_{k=1}^{N} M_{i k} \epsilon_{k}^{*}   \\
    Q_{i}^{A} & =\beta^{A} \sum_{k=1}^{N} M_{i k} \epsilon_{k}^{*} .
  \end{aligned}
  \label{eq:Q_approximation}
\end{equation}

Owing to the assumption that near the critical threshold \(p_{i}^{AI}(t), p_{i}^{UR}(t), p_{i}^{AR}(t) \rightarrow 0\) , we get \(p_i^A \approx p_i^{AS}\), \(p_i^{US} \approx 1-p_i^{AS} =1-p_i^A\). Therefore, Eq. (\ref{eq:mu2-epsilon_i}) can be rewritten as
\begin{equation}
  \mu_2 \epsilon_i^* = Q_i^A p_i^{A} + Q_i^U (1-p_i^{A}).
\end{equation}

Combining Eq. (\ref{eq:Q_approximation}) with Eq. (\ref{eq:mu2-epsilon_i}), it follows that
\begin{equation}
  \begin{aligned}
    \mu_{2} \epsilon_{i}^{*} & =\left[p_{i}^{A} \beta^{A}+\left(1-p_{i}^{A}\right) \beta^{U}\right] \sum_{k=1}^{N} M_{i k} \epsilon_{k}^{*} \\
                             & = \sum_{k=1}^{N}  \beta^{U}\left[1-(1-\sigma) p_{i}^{A}\right] M_{i k} \epsilon_{k}^{*} .
  \end{aligned}
\end{equation}

Let \(H_{i k}=\left[1-(1-\sigma) p_{i}^{A}\right] M_{i k}\), then we have
\begin{equation}
  \mu_2\epsilon_{i}^{*} = \beta^{U} \sum_{j=1}^{N} H_{i k} \epsilon_{k}^{*}.
\end{equation}
At this stage, the linearized dynamics of $\epsilon_i^*$ form a homogeneous system.
A non-trivial endemic solution ($\vec{\epsilon}\neq 0$) exists only when the associated linear operator
admits an eigenvalue larger than unity.
Thus, the epidemic threshold is determined by the condition that the largest eigenvalue
of the matrix $\mathbf{H}$ balances the recovery-to-infection ratio.
Accordingly, the equation can be written in eigenvalue form as
\begin{equation}
  \frac{\mu_2}{\beta^U} \vec\epsilon = \mathbf{H}\vec\epsilon.
\end{equation}

Thus, the epidemic threshold can be obtained by
\begin{equation}
  \beta_c=\frac{\mu_2}{\Lambda_{\text{max}}(\mathbf{H})}.
  \label{eq:beta_c}
\end{equation}

When the basic infection rate \( \beta \) is close to the epidemic threshold \( \beta_c \), the proportions of infected individuals are negligible, which refers to \( p_i^{AI}, p_i^{UR} \) and \( p_i^{AR} \rightarrow 0\), and the fact is that in the long run \(n_i(t) \approx n_i(0)\) when \(t \rightarrow \infty\). Although the awareness caused by disease-related information can hardly spread since the proportion of infected individuals is close to zero, there are still some population fluctuations due to the return decision mechanism.

In the following, we analyze the eigenvalues of the matrix \(\mathbf{H}\) under the assumption of non-zero mobility (\(g \neq 0\)) using perturbation theory. The matrix \(\mathbf{H}\) governs the coupled dynamics of disease and information propagation in the proposed metapopulation network. By decomposing \(\mathbf{H}\) into an unperturbed term and perturbative corrections, we can derive the eigenvalues in approximation according to the perturbation theory.

Expanding \(\mathbf{H}\) as a power series in \(g\), we express:
\begin{equation}
  \mathbf{H} = \mathbf{H}^{(0)} + g\mathbf{H}^{(1)} + g^2\mathbf{H}^{(2)} + \mathcal{O}(g^3),
\end{equation}
where \(\mathbf{H}^{(0)} = \text{diag}(\mathcal{L}_1 n_1, \mathcal{L}_2 n_2, \ldots, \mathcal{L}_N n_N)\) represents the unperturbed system with \(\mathcal{L}_i = 1 - (1-\sigma)p_i^A\). The first-order perturbation term \(g\mathbf{H}^{(1)}\), characterized by diagonal elements \(H_{ii}^{(1)} = -2\mathcal{L}_i \theta_i n_i\), captures the linear suppression of local infections due to population outflows. The second-order term \(g^2\mathbf{H}^{(2)}\), defined as \(H_{ik}^{(2)} = \mathcal{L}_i \theta_i \theta_k \sum_{j=1}^M R_{ij}R_{kj}n_k\), quantifies the nonlinear enhancement of inter-patch transmission mediated through transfer stations.

Hence, the eigenvalues of \(\mathbf{H}\) when \(g \ll 1\) are approximated as
\begin{equation}
  \lambda_i = \mathcal{L}_i n_i - 2g\mathcal{L}_i \theta_i n_i + g^2\mathcal{L}_i \theta_i^2 \sum_{j=1}^M R_{ij}^2 n_i + \mathcal{O}(g^3).
\end{equation}

Notably, the static scenario (\(g = 0\)) corresponds to a decoupled system where patches operate independently. In this limit, \(\mathbf{H}\) reduces to a diagonal matrix \(\mathbf{H}^{(0)}\), whose eigenvalues \(\lambda_i^{(0)} = \mathcal{L}_i n_i\) are trivially determined by the local parameters (\(\sigma, p_i^A, n_i\)). Since no inter-patch coupling exists, perturbation theory becomes redundant as the exact solutions are already accessible without higher-order corrections.

\begin{figure*}[ht!]
  \centering
  \hspace*{-0.9em} 
  \begin{subfigure}[b]{0.35\textwidth}
    \centering
    \includegraphics[width=\textwidth]{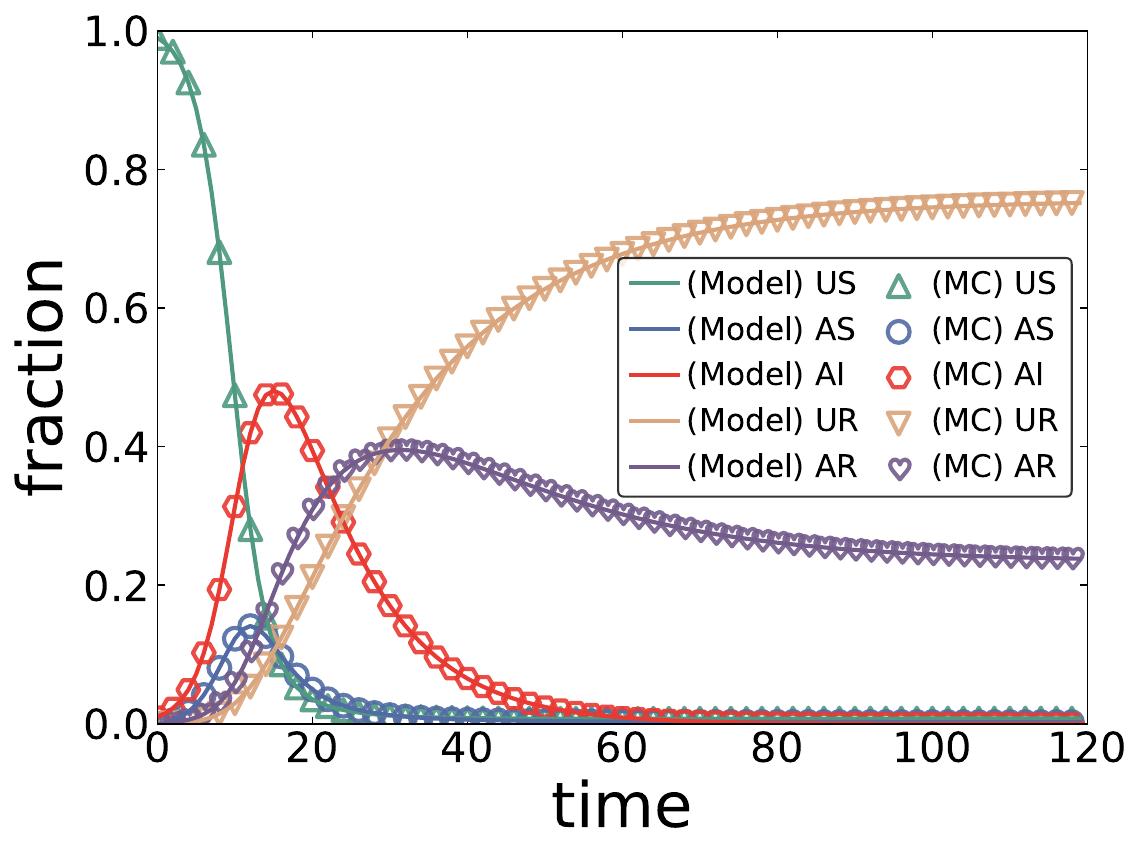}
    \caption{}
    \label{test1_1}
  \end{subfigure}
  \hspace*{-0.9em} 
  \begin{subfigure}[b]{0.335\textwidth}
    \centering
    \includegraphics[width=\textwidth]{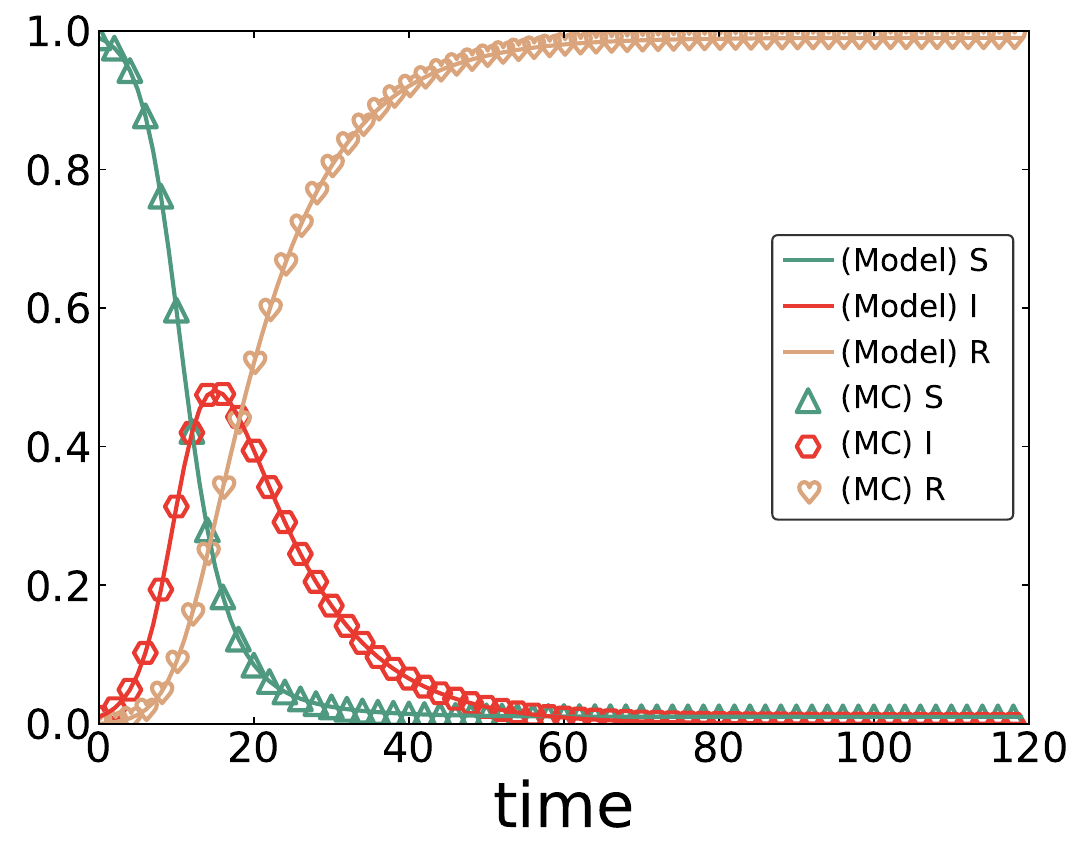}
    \caption{}
    \label{test1_2}
  \end{subfigure}
  \hspace*{-0.9em} 
  \begin{subfigure}[b]{0.335\textwidth}
    \centering
    \includegraphics[width=\textwidth]{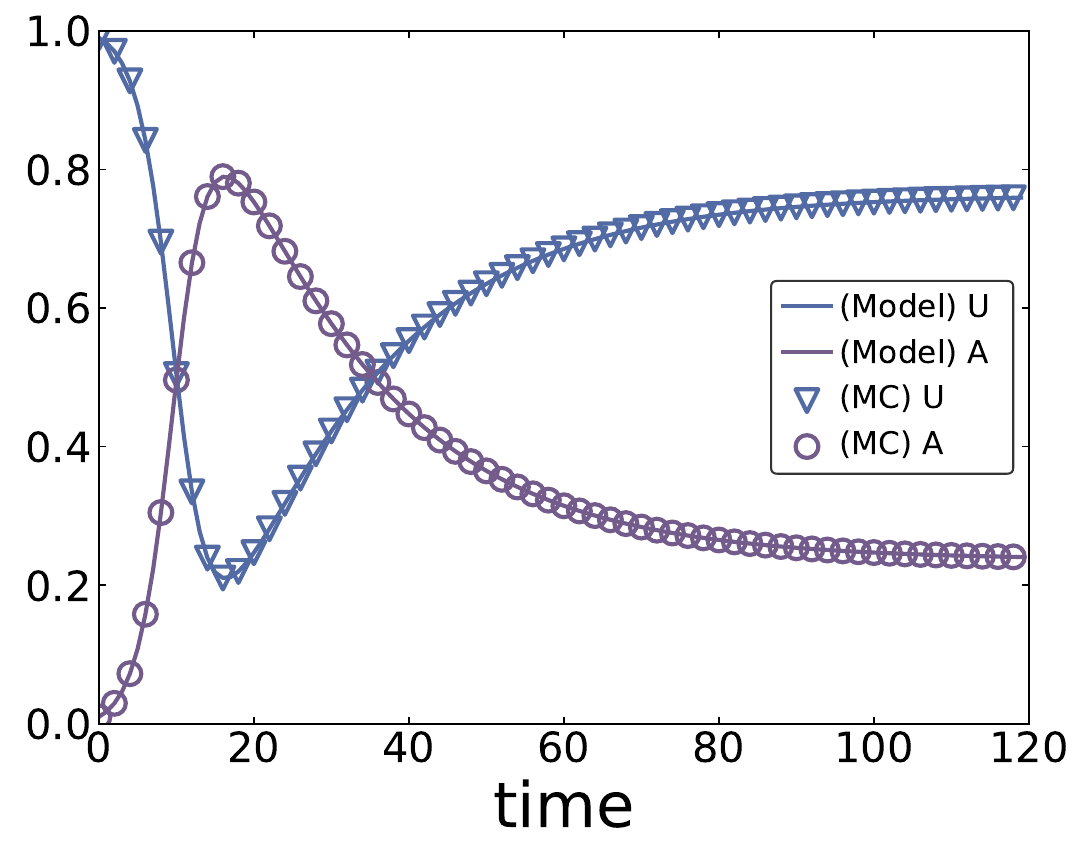}
    \caption{}
    \label{test1_3}
  \end{subfigure}
  \caption{
    \textbf{Comparison between theoretical predictions from the MMCA equations and Monte Carlo simulations over 120 time steps. } Panel (a) shows the time evolution of five global compartmental states (US, AS, AI, UR, AR) in the coupled UAU–SIR system.
    Panel (b) shows the physical SIR layer including susceptible (S), infected (I), and recovered (R) states.
    Panel (c) shows the cyber-layer UAU dynamics, capturing the evolution of unaware (U) and aware (A) individuals.
    The x-axis indicates discrete time steps and the y-axis indicates the population fraction.
    Solid curves represent results computed from MMCA equations and symbols represent Monte Carlo simulations averaged over 200 realizations.
  }
  \label{fig_1}
\end{figure*}

\section{NUMERICAL SIMULATION}
\label{sec4:numerical_simulation}
In this section, we conduct comprehensive numerical simulations to validate the theoretical framework and examine the impacts of information diffusion and event-triggered migration on epidemic dynamics within a metapopulation networked CPS.
The simulations are organized into five parts:
1) validation of the theoretical predictions;
2) analysis of model dynamics under diverse parameter combinations;
3) investigation of coupled influence of information and mobility;
4) evaluation of the rendezvous effect under heterogeneous conditions;
5) benchmarking with classical metapopulation and awareness models.
Unless otherwise specified, the default parameters are set as: initial infection ratio = 1\% randomly selected in residences, \(N = 15\), \(M = 5\), \(\beta = 0.003\), \(\lambda = 0.2\), \(\mu_1 = 0.15\), \(\mu_2 = 0.1\), \(g = 0.5\), \(\varepsilon_0 = 0.2\), \(\alpha = 0.3\), \(\sigma = 0.5\), \(\theta = 1\), and \(n_i(0) = 200\). Each point calculated by the MC simulations is averaged over 200 independent realizations.

\subsection{Validation of Theoretical Predictions}

To validate the effectiveness of the proposed model in capturing epidemic dynamics on metapopulation networks, we compare theoretical predictions derived from the MMCA equations with the results obtained via Monte Carlo (MC) simulations. Based on Eqs. (\ref{eq:single_states}) and (\ref{eq:MMCA}), we present the temporal evolution of epidemic compartmental states across the entire system and in the single layers of SIR and UAU dynamics over 120 time steps.

As shown in Fig. \ref{fig_1}, the solid curves correspond to the results computed using MMCA equations, while the symbols represent the averaged MC simulations. The subplots (a), (b), and (c) illustrate the evolution of compartmental fractions in the networked CPS, physical layer, and virtual layer, respectively.
In Fig. \ref{fig_1}(a), the early phase of the dynamics is dominated by the US and AI states. As time progresses, the fraction of US individuals declines sharply and vanishes after approximately 20 time steps, indicating rapid information dissemination and infection onset. Simultaneously, the fraction of AI states first increases and then drops, peaking around \(t = 15\), corresponding to the epidemic outbreak phase. Eventually, the system stabilizes into a disease-free equilibrium, where individuals predominantly occupy the UR and AR  states.
Fig.~\ref{fig_1}(b) highlights the SIR dynamics within the physical layer, while Fig.~\ref{fig_1}(c) illustrates the awareness evolution within the virtual layer and the system ultimately stabilizes with approximately 80\% of the population remaining in the aware state, reflecting a long-term memory of the awareness process.
The high consistency between MMCA theoretical predictions and MC simulations across all three panels demonstrates the robustness and reliability of the proposed model in capturing both information progression and epidemic dynamics in a coupled metapopulation networked CPS.

\subsection{Analysis of Model Dynamics under Diverse Parameter Combinations}

\begin{figure*}[ht!]
  \centering
  \hspace*{-0.9em}
  \begin{subfigure}[b]{0.34\textwidth}
    \centering
    \includegraphics[width=\textwidth]{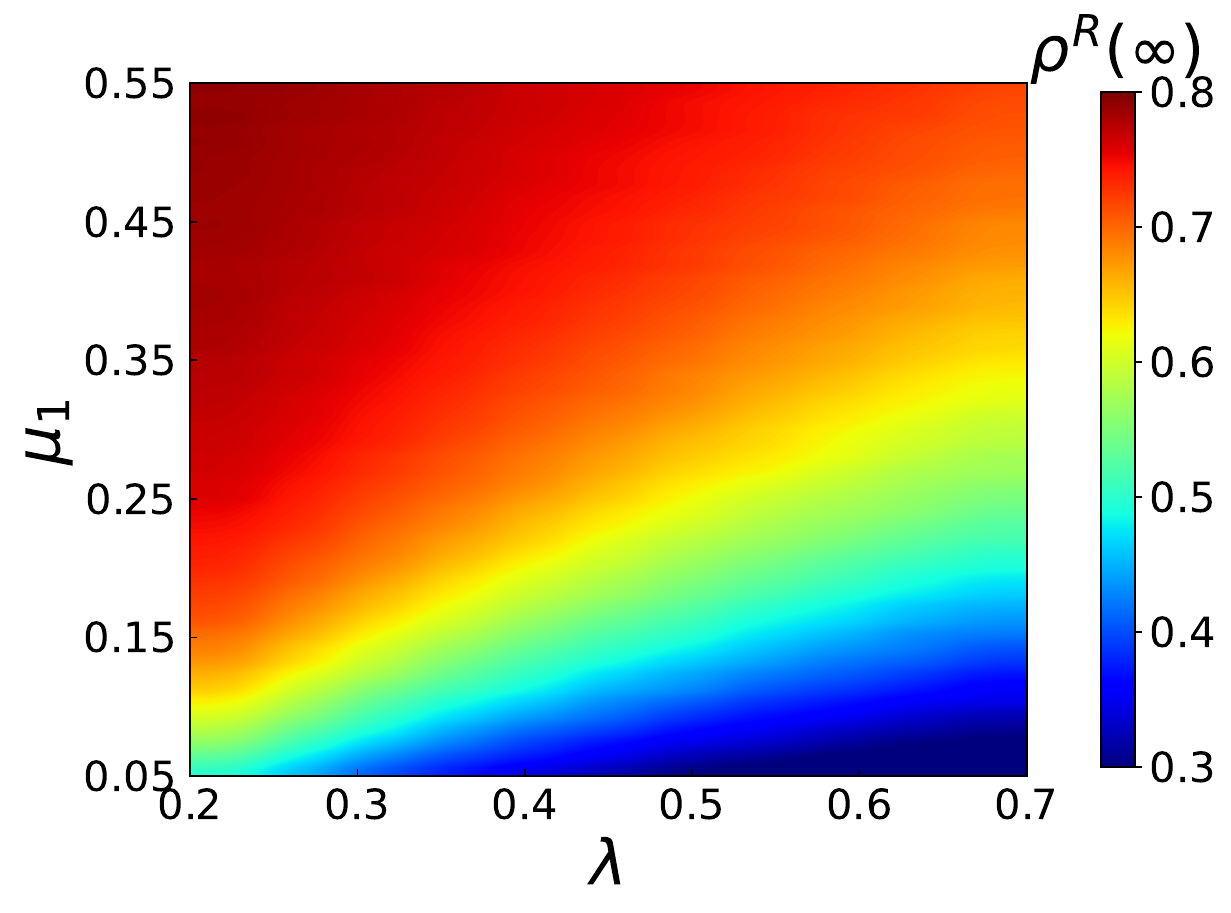}
    \caption{}
    \label{test2_1}
  \end{subfigure}
  \hspace*{-0.8em}
  \begin{subfigure}[b]{0.34\textwidth}
    \centering
    \includegraphics[width=\textwidth]{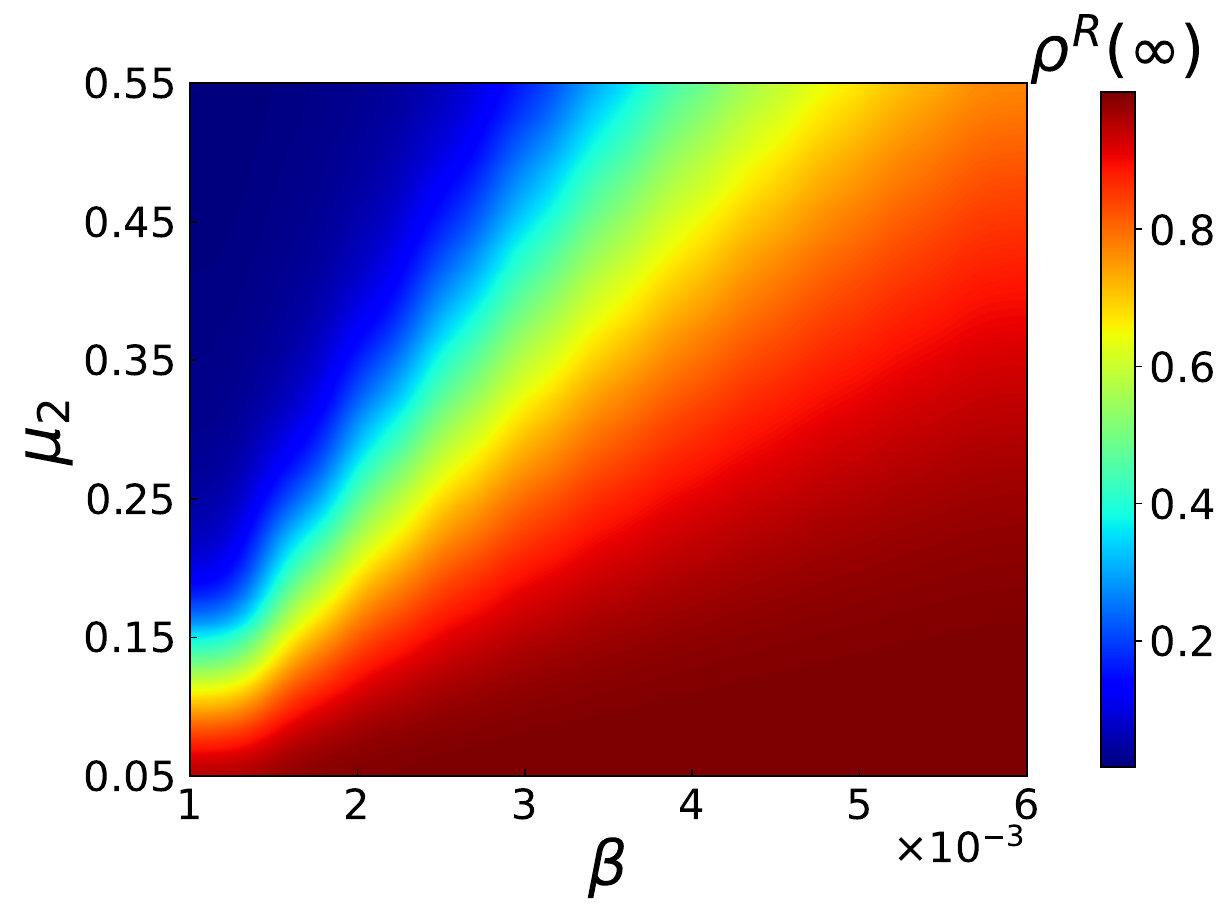}
    \caption{}
    \label{test2_2}
  \end{subfigure}
  \hspace*{-0.8em}
  \begin{subfigure}[b]{0.33\textwidth}
    \centering
    \includegraphics[width=\textwidth]{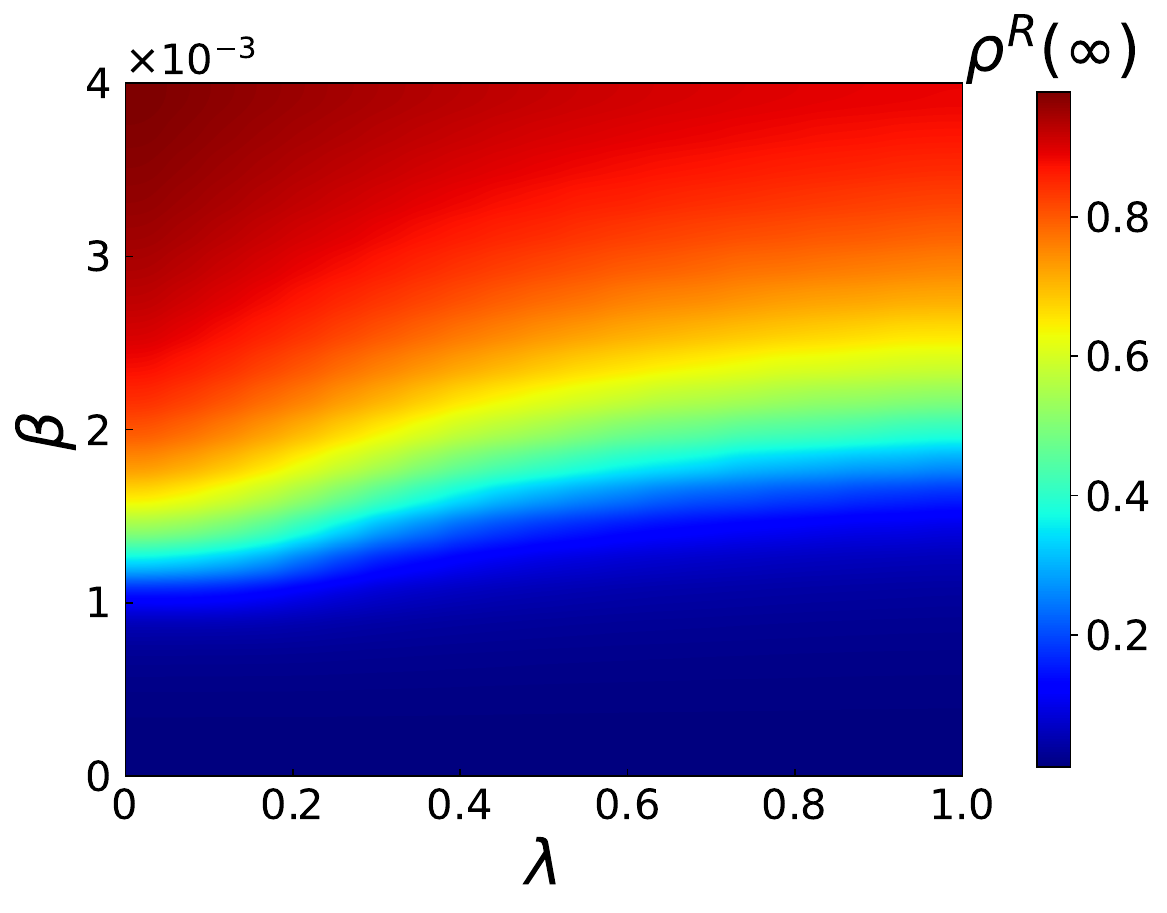}
    \caption{}
    \label{test2_3}
  \end{subfigure}
  \caption{
    \textbf{Final epidemic size \(\rho^R(\infty)\) under different parameter combinations over 200 time steps.} Panel (a) shows $\rho^R(\infty)$ as a function of information transmission rate $\lambda$ and information forgetting rate $\mu_1$, with $\lambda$ from 0.2 to 0.7 and $\mu_1$ from 0.05 to 0.55.
    Panel (b) shows $\rho^R(\infty)$ as a function of disease spreading rate $\beta$ between 0.001 and 0.006 and recovery rate $\mu_2$ between 0.05 and 0.55.
    Panel (c) shows $\rho^R(\infty)$ in the $\lambda$–$\beta$ space.
    Color gradients indicate the proportion of final recovered individuals, as denoted by the color bars.
    Each value is averaged over 200 Monte Carlo realizations.
  }
  \label{fig_2}
\end{figure*}

Aiming at exploring the impact of different parameters on the epidemic dynamics, we define the final recovered fraction \(\rho^R(\infty)\) as the primary indicator of epidemic size. In the context of the SIR model, where all infected individuals eventually recover, \(\rho^R(\infty)\) also serves as the cumulative infected proportion.

Fig.~\ref{fig_2} presents the values of \(\rho^R(\infty)\) in the steady state as functions of different combinations of epidemic and awareness-related parameters, evaluated over 200 simulation steps.
In Fig. \ref{fig_2}(a), the heatmap depicts the variation of \(\rho^R(\infty)\) with respect to the information transmission rate \(\lambda\) (ranging from 0.2 to 0.7) and the information forgetting rate \(\mu_1\) (ranging from 0.05 to 0.55). A clear gradient is observed, where increased \(\lambda\) and reduced \(\mu_1\) jointly result in significantly lower values of \(\rho^R(\infty)\). This trend highlights the critical role of persistent awareness in mitigating the epidemic spread, as more efficient and sustained information dissemination leads to a better containment.
Fig.~\ref{fig_2}(b) illustrates the dependence of \(\rho^R(\infty)\) on the disease spreading rate \(\beta\) and the recovery rate \(\mu_2\). A distinct transition can be observed, consistent with the outbreak threshold predicted by Eq. (\ref{eq:beta_c}). When \(\beta\) surpasses the critical value (approximately 0.0012), the system shifts abruptly from a disease-free state to a widespread outbreak, resembling a discontinuous phase transition. This transition becomes sharper at lower recovery rates \(\mu_2\), indicating the vulnerability of the system to slow recovery processes.
Fig.~\ref{fig_2}(c) investigates the interplay between information transmission (\(\lambda\)) and disease spreading (\(\beta\)) rates. For fixed values of \(\beta\), an increase in \(\lambda\) leads to a marked decrease in the final infection level, suggesting that effective information dissemination can elevate the epidemic threshold and delay the onset of outbreaks. Notably, at low \(\lambda\), even moderate values of \(\beta\) can trigger widespread epidemics.

\subsection{Investigation of Coupled Influence of Information and Mobility}

\begin{figure*}[ht!]
  \centering
  \begin{subfigure}[b]{0.46\textwidth}
    \centering
    \includegraphics[width=\textwidth]{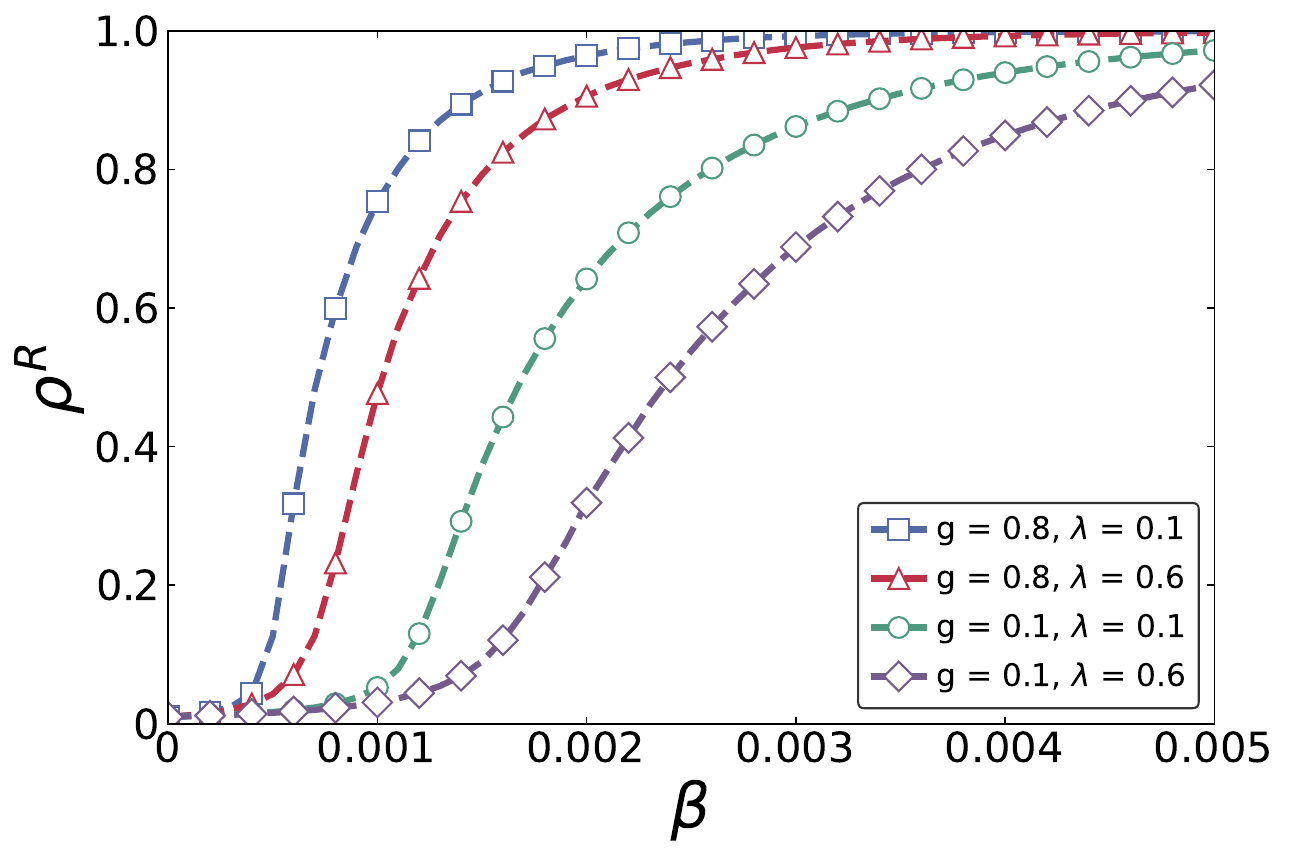}
    \caption{}
    \label{test3_1}
  \end{subfigure}
  \begin{subfigure}[b]{0.46\textwidth}
    \centering
    \includegraphics[width=\textwidth]{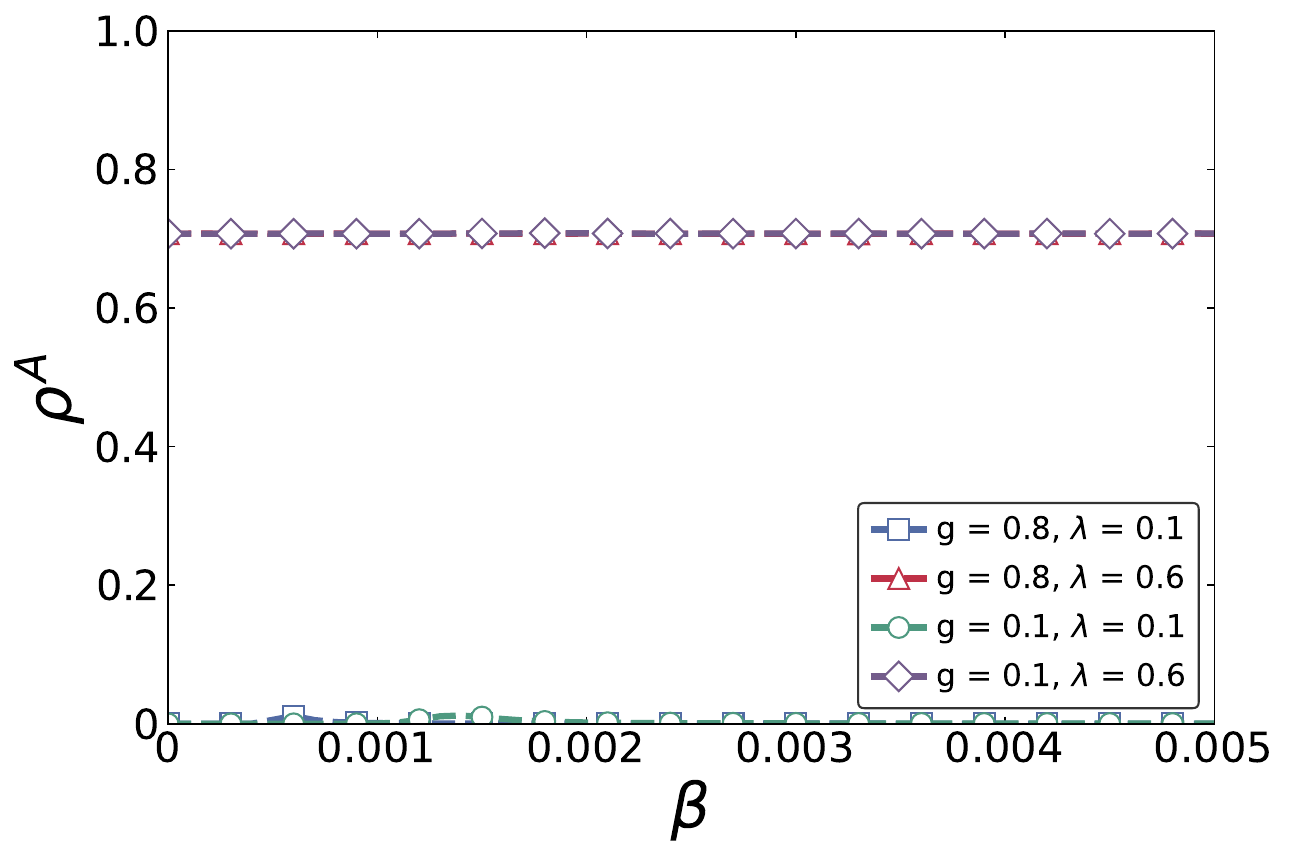}
    \caption{}
    \label{test3_2}
  \end{subfigure}
  \caption{
    \textbf{Influence of information diffusion and human mobility on epidemic dynamics.} Panels (a) and (b) plot the final recovered fraction $\rho^R(\infty)$ and final awareness level $\rho^A(\infty)$ as functions of the infection rate $\beta$.
    Both panels compare four parameter combinations: $\{g = 0.1, \lambda = 0.1\}$, $\{g = 0.1, \lambda = 0.6\}$, $\{g = 0.8, \lambda = 0.1\}$, and $\{g = 0.8, \lambda = 0.6\}$.
    The x-axis shows the infection rate $\beta$ and the y-axis shows the final fraction for either recovered (left) or aware (right) states.
    The Curves denote theoretical results and symbols denote Monte Carlo simulations averaged over 200 runs.
  }
  \label{fig_3}
\end{figure*}

\begin{figure*}[ht!]
  \centering
  \hspace*{-0.9em} 
  \begin{subfigure}[b]{0.35\textwidth}
    \centering
    \includegraphics[width=\textwidth]{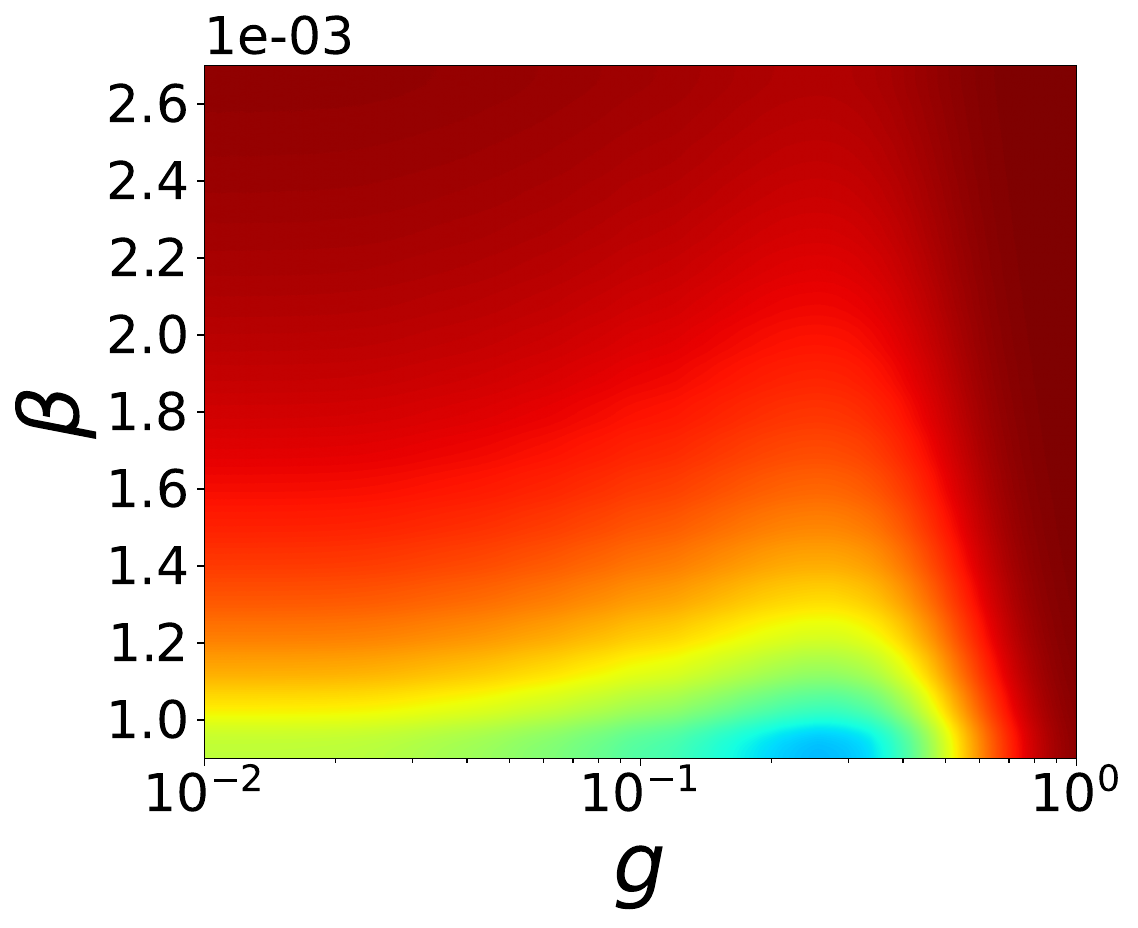}
    \caption{}
    \label{test4_1}
  \end{subfigure}
  \hspace*{-0.8em} 
  \begin{subfigure}[b]{0.32\textwidth}
    \centering
    \includegraphics[width=\textwidth]{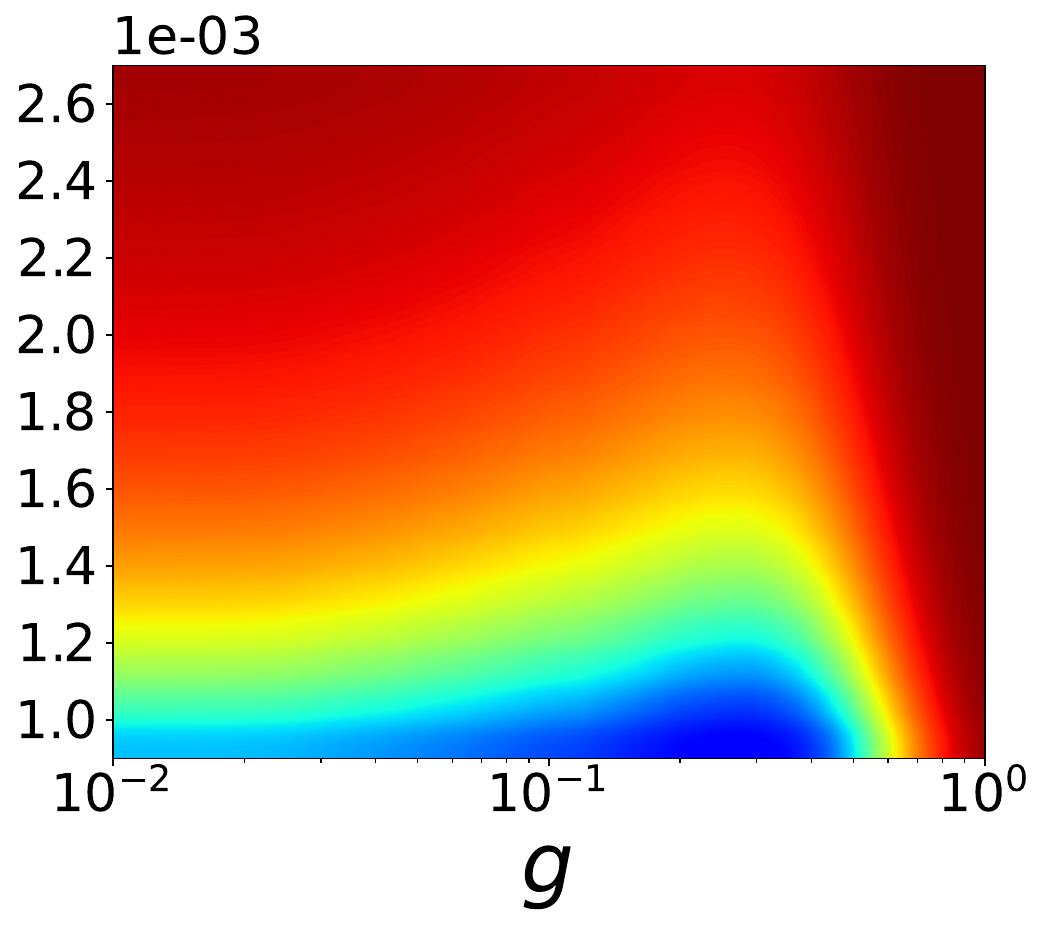}
    \caption{}
    \label{test4_2}
  \end{subfigure}
  \hspace*{-0.9em} 
  \begin{subfigure}[b]{0.35\textwidth}
    \centering
    \includegraphics[width=\textwidth]{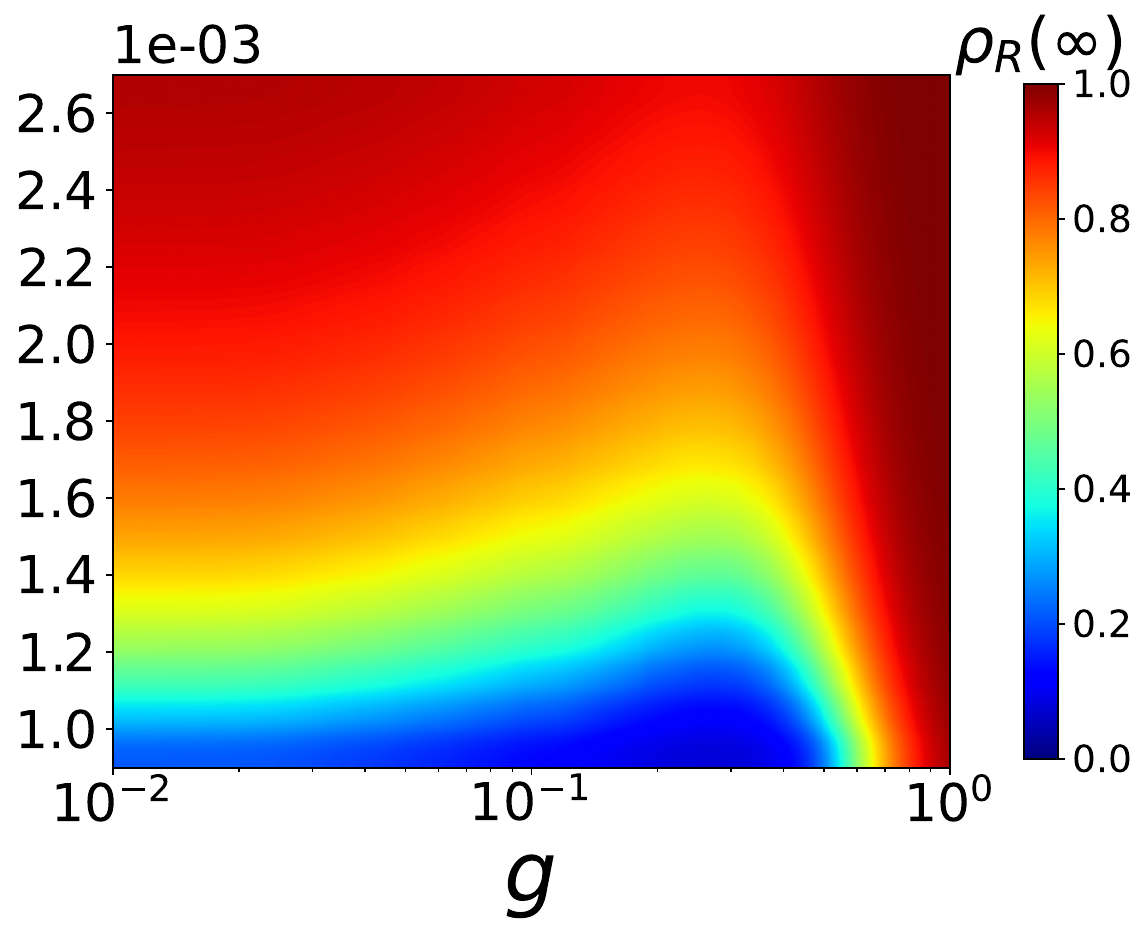}
    \caption{}
    \label{test4_3}
  \end{subfigure}
  \caption{
    \textbf{Total fraction of recovered individuals $\rho^R(\infty)$ in the steady state as a function of the disease spreading rate $\beta$ and mobility rate $g$, for three values of the information transmission rate: (a) $\lambda=0.2$, (b) $\lambda=0.5$, and (c) $\lambda=0.8$.}
    Panels (a) to (c) display heatmaps of  $\rho^R(\infty)$ across $g \in [0.01, 1]$ and $\beta$.
    The x-axis uses a logarithmic scale for the mobility rate $g$ and the y-axis represents the disease spreading rate $\beta$.
    All subplots share the same parameter settings and color scale, indicating final epidemic size from 0 (blue) to 1 (red).
  }
  \label{fig_4}
\end{figure*}

Next, we explore the coupled role of information diffusion and human mobility in shaping epidemic outcomes.
Fig.~\ref{fig_3} illustrates the steady-state behaviors of \(\rho^R(\infty)\) and \(\rho^A(\infty)\) under various combinations of mobility rate \(g\), information transmission rate \(\lambda\), and infection rate \(\beta\). In Fig.~\ref{fig_3}(a), the curves show that increasing mobility rate \(g\) leads to a more rapid rise in \(\rho^R(\infty)\) as \(\beta\) increases. For example, the blue square and red triangle curves (\(g = 0.8\)) reach saturation at significantly lower \(\beta\) values compared to the green circle and purple diamond curves (\(g = 0.1\)), indicating that elevated mobility enhances contact opportunities and accelerates disease transmission.
At fixed mobility (\(g = 0.1\)), a higher information transmission rate (\(\lambda = 0.6\), purple diamonds) substantially increases the epidemic threshold compared to the lower transmission rate case (\(\lambda = 0.1\), green circles), thereby delaying the outbreak onset. This demonstrates that information dissemination plays a critical role in flattening the outbreak curve and elevating the epidemic threshold.
Fig.~\ref{fig_3}(b) shows the corresponding steady-state fraction of aware individuals \(\rho^A(\infty)\) for each parameter setting. The curves remain flat and invariant with respect to \(\beta\), reflecting that information dynamics evolve independently of disease progression. Notably, for \(\lambda = 0.6\), the system maintains a high awareness level (approximately 0.75) regardless of mobility rate or infection level. In contrast, when \(\lambda = 0.1\), the awareness fraction remains near zero—even under high mobility, highlighting that insufficient information dissemination fundamentally limits the spread of preventive behavior.

Fig.~\ref{fig_4} further visualizes the final epidemic size \(\rho^R(\infty)\) across a two-dimensional parameter space defined by the mobility rate \(g\) and infection rate \(\beta\), under different levels of awareness diffusion.
As \(\lambda\) increases, a clear pattern emerges in Fig.~\ref{fig_4}(c): the blue low-prevalence regions expand markedly, indicating that a high rate of information transmission significantly suppresses epidemic outbreaks. This shift suggests that effective awareness diffusion can enhance the resilience of the system to infection, even under varying levels of mobility.
Importantly, we adopt a logarithmic scale on the \(x\)-axis (mobility rate \(g\)) to highlight the non-monotonic influence of mobility. The boundary gradients in all panels suggest that mobility does not uniformly amplify contagion. Specifically, within the low mobility range (\(g \in [0.01, 0.3]\)), moderate increases in mobility may actually inhibit disease propagation. This counterintuitive behavior suggests that restricting migrant population mobility to low levels helps curb viral spread.

\subsection{Evaluation of the Rendezvous Effect under Heterogeneous Conditions}

\begin{figure*}[ht!]
  \centering
  \hspace*{-0.8em} 
  \begin{subfigure}[b]{0.34\textwidth}
    \centering
    \includegraphics[width=\textwidth]{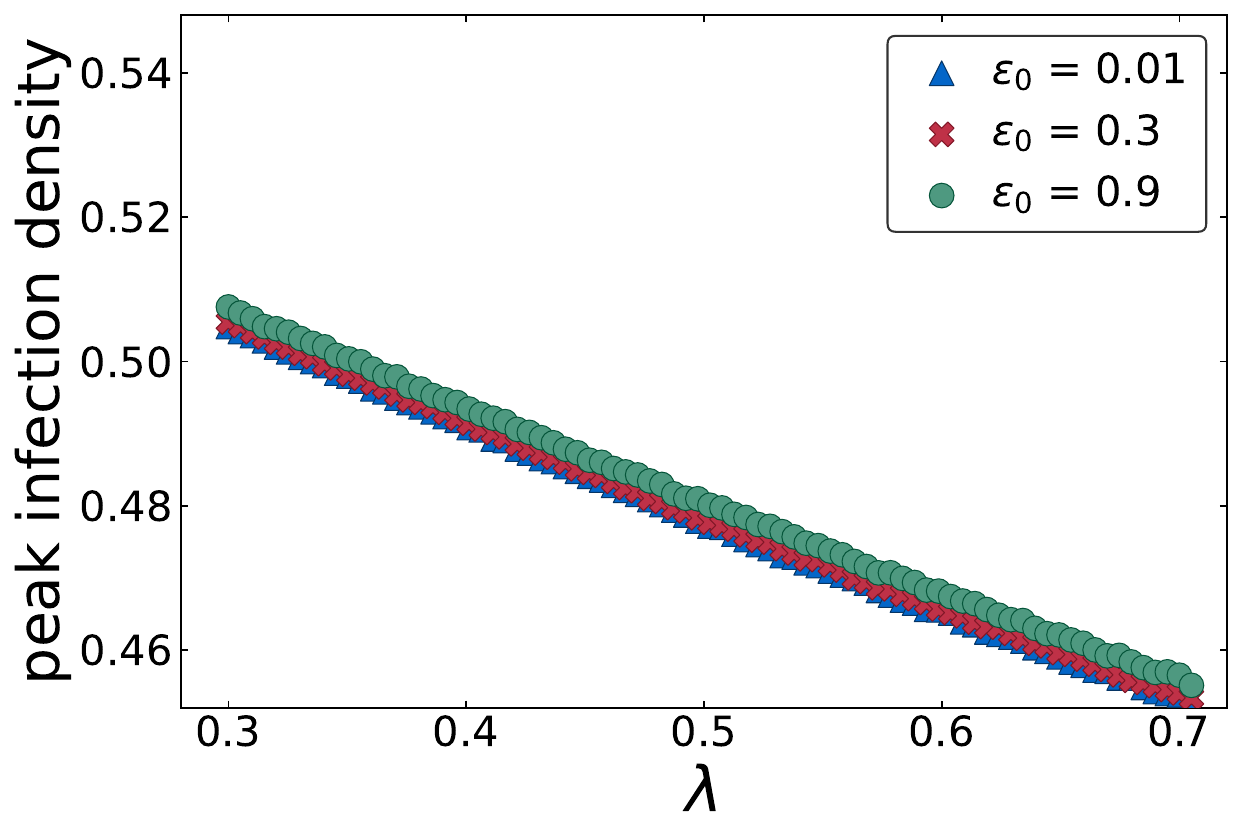}
    \caption{}
    \label{test5_1}
  \end{subfigure}
  \hspace*{-0.8em} 
  \begin{subfigure}[b]{0.323\textwidth}
    \centering
    \includegraphics[width=\textwidth]{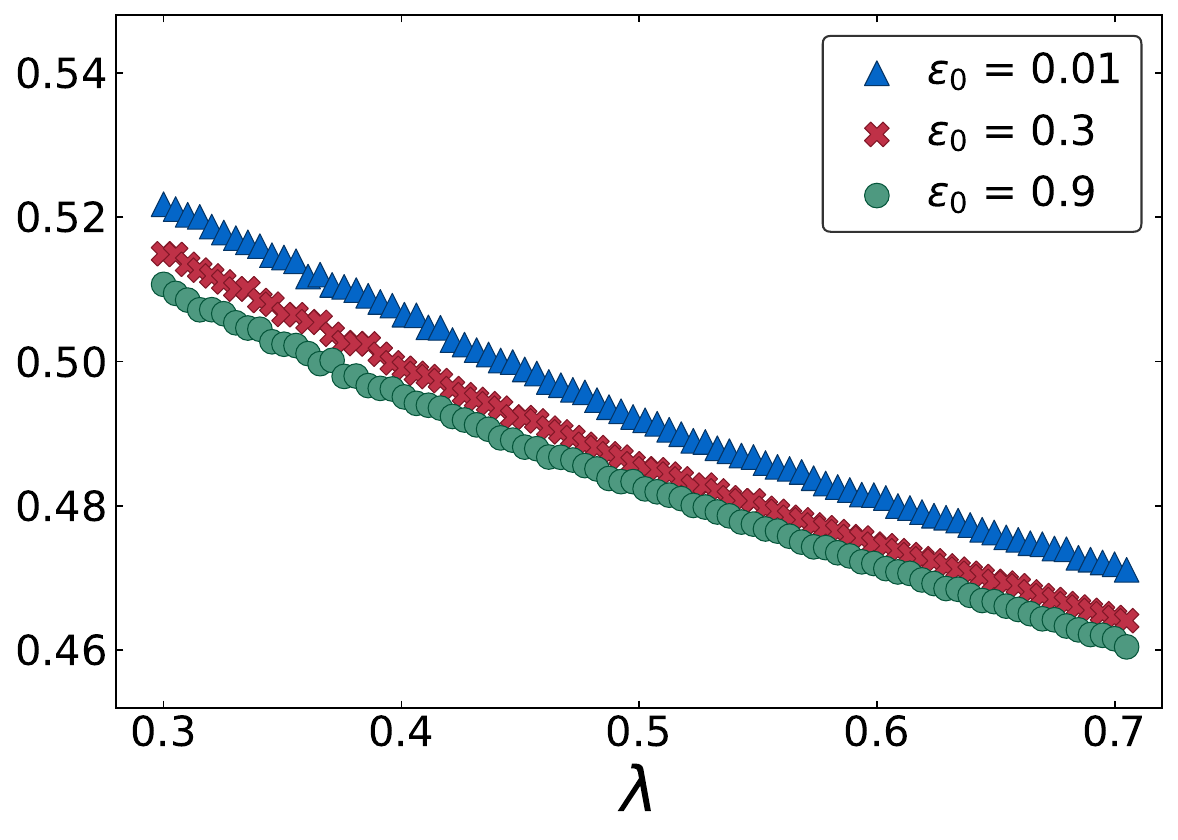}
    \caption{}
    \label{test5_2}
  \end{subfigure}
  \hspace*{-0.8em} 
  \begin{subfigure}[b]{0.323\textwidth}
    \centering
    \includegraphics[width=\textwidth]{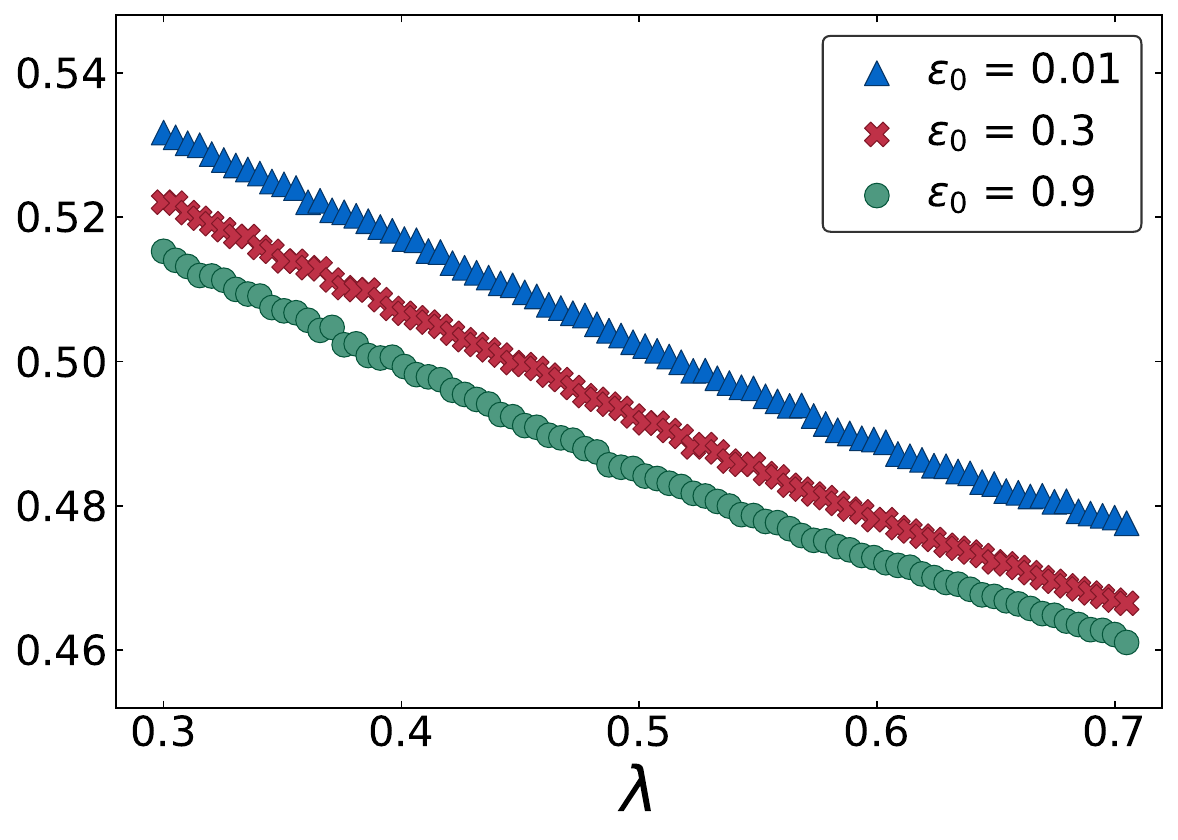}
    \caption{}
    \label{test5_3}
  \end{subfigure}
  \caption{
    \textbf{Peak infection densities under varying activation intensity \(\varepsilon_0\) across different types of initial population configurations.}
    Panel (a) uses uniform distribution $n_i(0) = 450$, panel (b) uses moderately heterogeneous distribution $n_i(0) = 50i + 50$, and panel (c) uses highly heterogeneous distribution $n_i(0) = 60i - 30 $, where $i = 1, 2, ..., N$.
    The x-axis shows the value of $\varepsilon_0$ and the y-axis shows the peak infected proportion observed during simulation.
    Each case involves 20 patches (15 residences and 5 transfer stations), totaling \(6.75 \times 10^3\) individuals.
  }
  \label{fig_5}
\end{figure*}

To systematically examine the interplay between event-triggered migration and spatial heterogeneity, we assess how population structure and behavioral parameters affect the peak infection density. Our simulations focus on three key aspects: 1) the sensitivity of epidemic peaks to activation intensity under different levels of spatial heterogeneity; 2) the combined effects of activation threshold and intensity on disease dynamics; and 3) the role of rendezvous nodes (transfer stations), in shaping the overall magnitude of outbreaks.

Fig.~\ref{fig_5} presents the peak infection densities under different activation intensities \(\varepsilon_0\), across different levels of population heterogeneity.
In the uniform case (panel a), the curves for varying \(\varepsilon_0\) are nearly parallel, indicating negligible influence of migration on infection dynamics due to spatial homogeneity.
This suppression effect is attributed to enhanced population dispersal once awareness is triggered, which alleviates infection clustering in densely populated patches. The results highlight the importance of adaptive migration in mitigating epidemic severity under imbalanced initial conditions.

\begin{figure*}[ht!]
  \centering
  \begin{subfigure}[b]{0.46\textwidth}
    \centering
    \includegraphics[width=\textwidth]{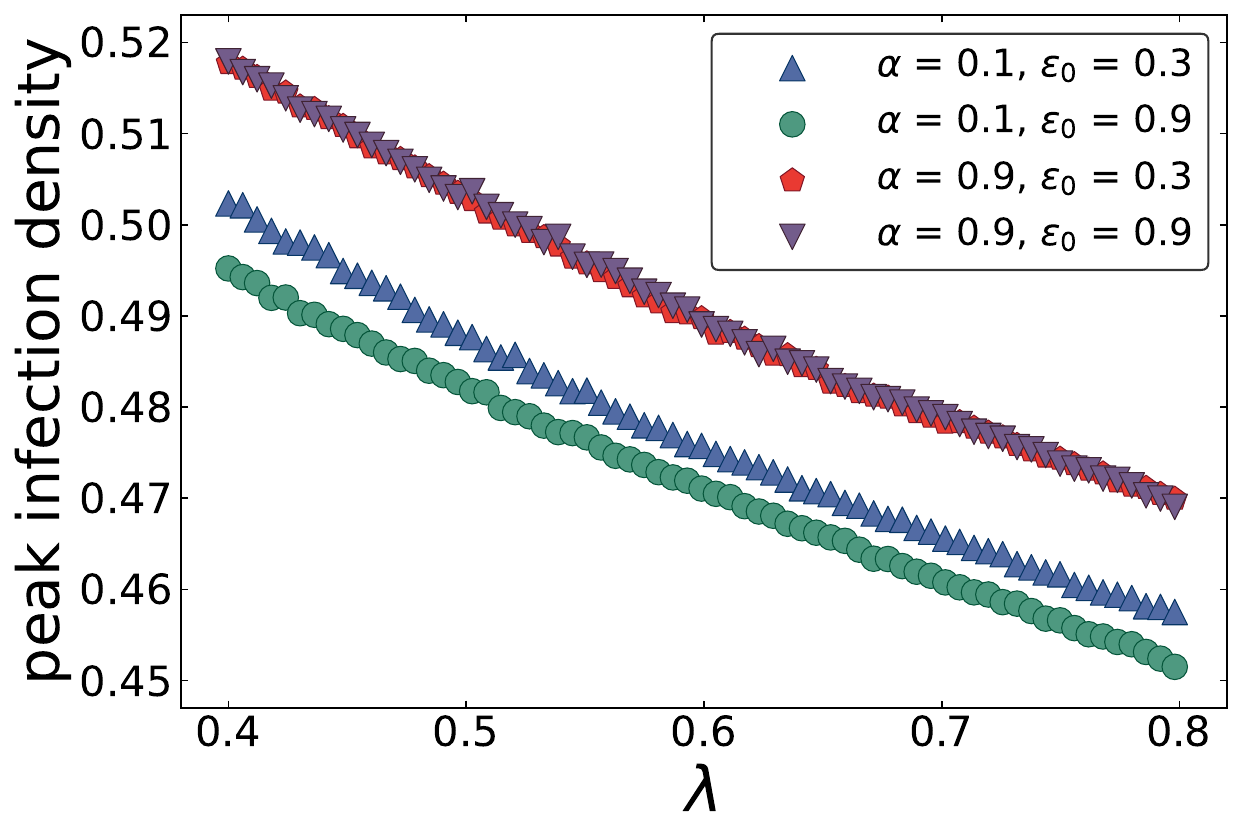}
    \caption{}
    \label{test6_1}
  \end{subfigure}
  \hspace*{0.4em} 
  \begin{subfigure}[b]{0.435\textwidth}
    \centering
    \includegraphics[width=\textwidth]{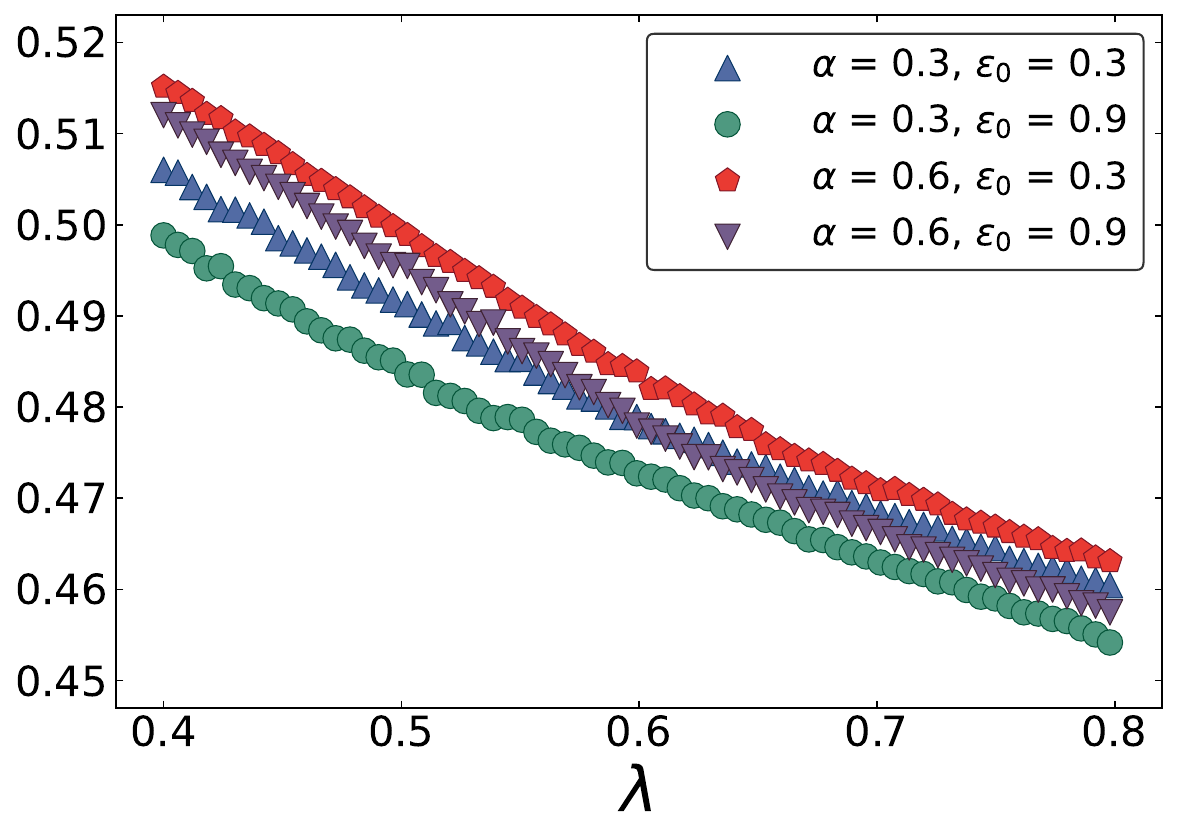}
    \caption{}
    \label{test6_2}
  \end{subfigure}
  \caption{
    \textbf{Peak infection densities as a function of information transmission rate \(\lambda\) for different combinations of activation threshold \(\alpha\) and intensity \(\varepsilon_0\). }
    Panel (a) corresponds to $\alpha = 0.1$ and \(0.9\), and panel (b) to $\alpha = 0.3$ and \(0.6\) with $\varepsilon_0 = 0.3$ and $0.9$ in each case.
    The x-axis shows $\lambda \in [0.4, 0.8]$ and the y-axis shows the corresponding peak infection density.
    All cases assume heterogeneous initial populations as $n_i(0) = 60i - 30 $, where $i = 1, 2, ..., N$.
    Each line corresponds to a unique parameter setting and is averaged over 200 independent simulations.
  }
  \label{fig_6}
\end{figure*}

To further investigate the behavioral dynamics, we explore how the activation threshold \(\alpha\) and intensity \(\varepsilon_0\) jointly influence the epidemic response.
Fig.~\ref{fig_6} shows the peak infection density as a function of the information transmission rate \(\lambda\), under different configurations of \(\alpha\) and \(\varepsilon_0\).
As illustrated, lower values of \(\alpha\) consistently reduce peak infection levels, particularly when combined with large \(\varepsilon_0\). This suggests that a lower activation threshold for migration allows early population flow to inhibit disease concentration.
In Fig.~\ref{fig_6}(\subref{test6_1}), the gap between \(\varepsilon_0 = 0.3\) and \(\varepsilon_0 = 0.9\) curves remains small when \(\alpha = 0.9\), indicating that high awareness thresholds delay migration until the infection peak is approached. Conversely, Fig.~\ref{fig_6}(\subref{test6_2}) reveals that when \(\alpha\) is set to 0.3 or 0.6, low thresholds prompt early migration and substantially reduce infection peaks—especially under high \(\varepsilon_0\). A crossover between curves at moderate \(\lambda\) suggests that in early phases, threshold \(\alpha\) is the dominant factor, while in later phases, activation intensity \(\varepsilon_0\) determines the extent of mitigation. These results illustrate the nonlinear coupling between awareness dynamics and behaviorally adaptive mobility.

\begin{figure*}[ht!]
  \centering
  \begin{subfigure}[b]{0.46\textwidth}
    \centering
    \includegraphics[width=\textwidth]{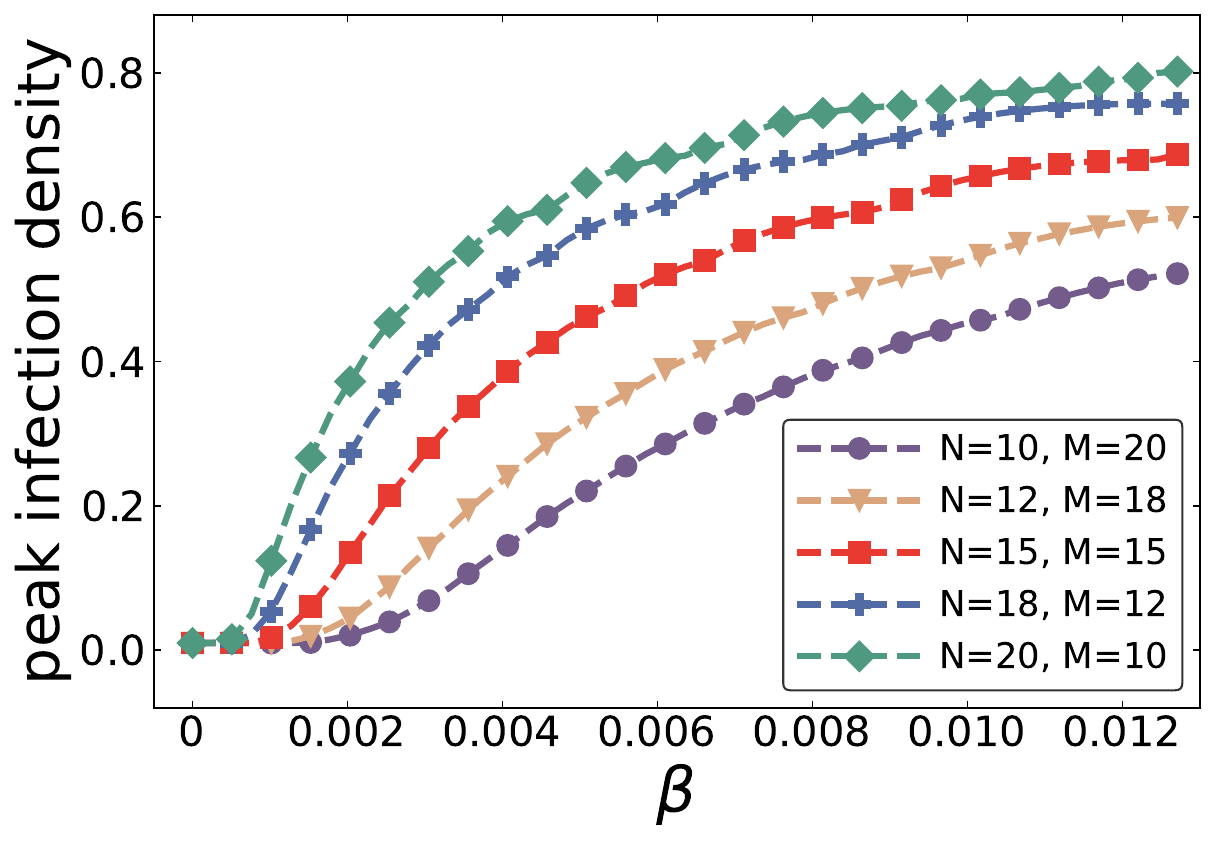}
    \caption{}
    \label{test7_1}
  \end{subfigure}
  \hspace*{0.4em} 
  \begin{subfigure}[b]{0.435\textwidth}
    \centering
    \includegraphics[width=\textwidth]{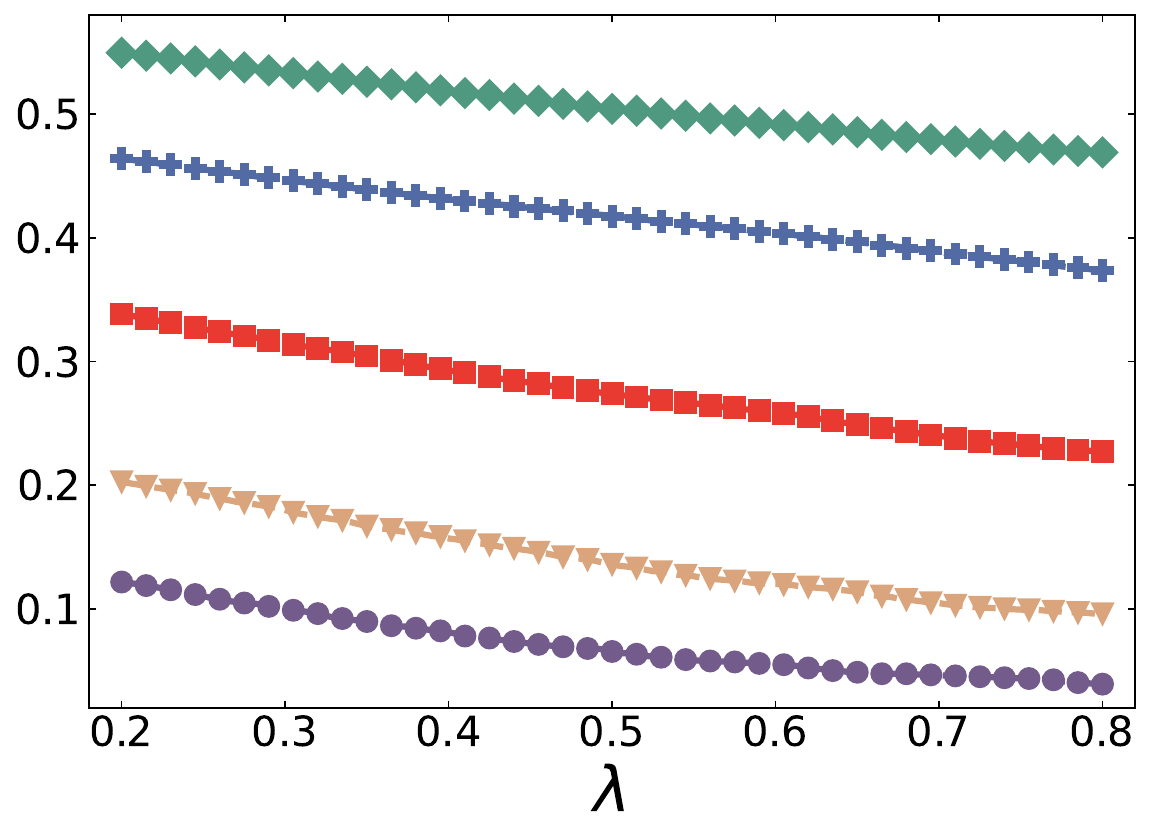}
    \caption{}
    \label{test7_2}
  \end{subfigure}
  \caption{
    \textbf{Peak infection densities for different configurations of residences (\(N\)) and transfer stations (\(M\)).}
    In panel (a), the x-axis represents $\beta$ from 0 to 0.012, and the y-axis shows peak infection density.
    In panel (b), the x-axis represents $\lambda$ between 0.2 and 0.8.   All configurations satisfy $N + M = 30$, with total population distributed as $n_i(0) = 60i - 30 $, where $i = 1, 2, ..., N$.
  }
  \label{fig_7}
\end{figure*}

\begin{figure*}[ht!]
  \centering
  \hspace*{-0.9em}
  \begin{subfigure}[b]{0.335\textwidth}
    \centering
    \includegraphics[width=\textwidth]{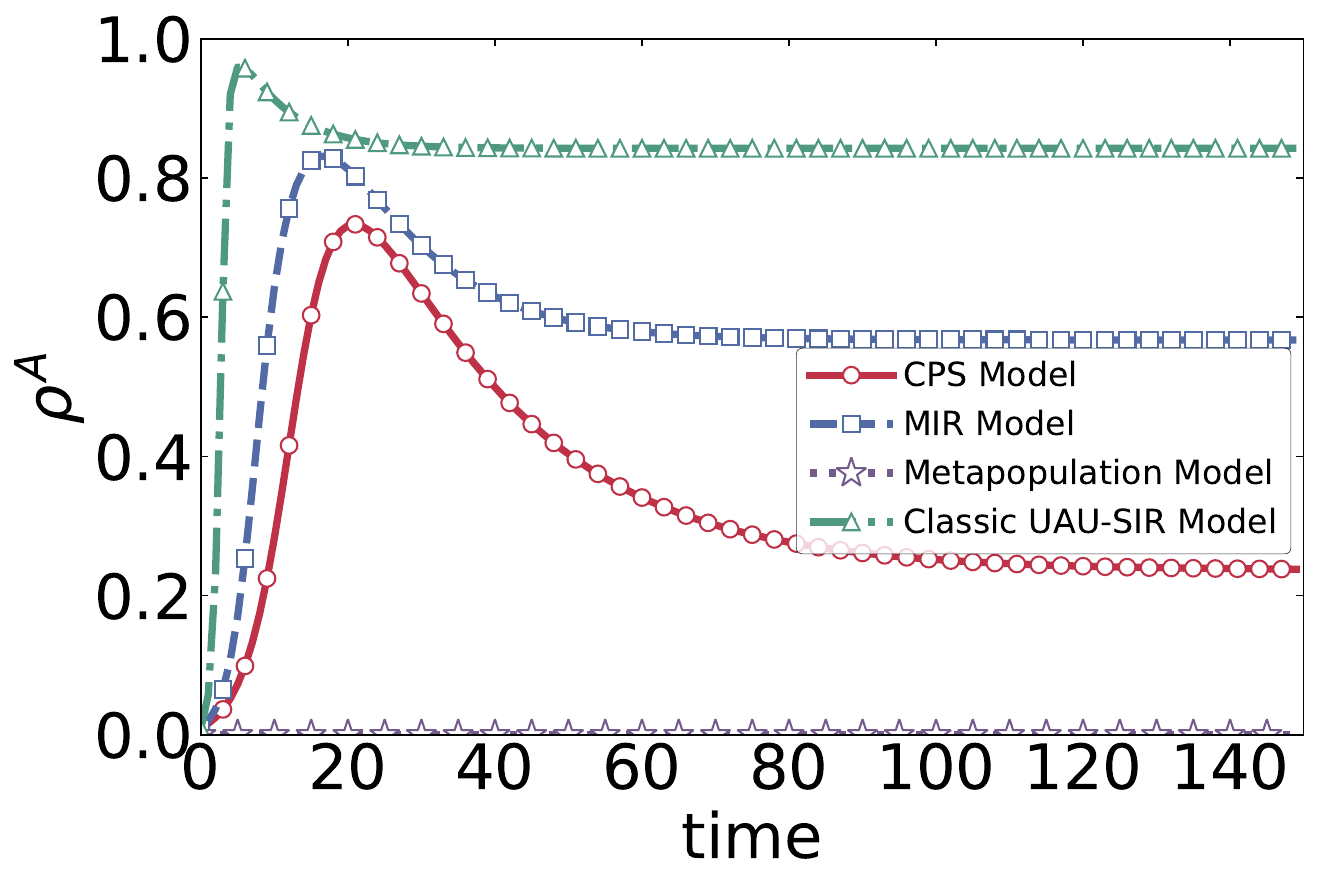}
    \caption{}
    \label{test8_1}
  \end{subfigure}
  \hspace*{-0.8em}
  \begin{subfigure}[b]{0.335\textwidth}
    \centering
    \includegraphics[width=\textwidth]{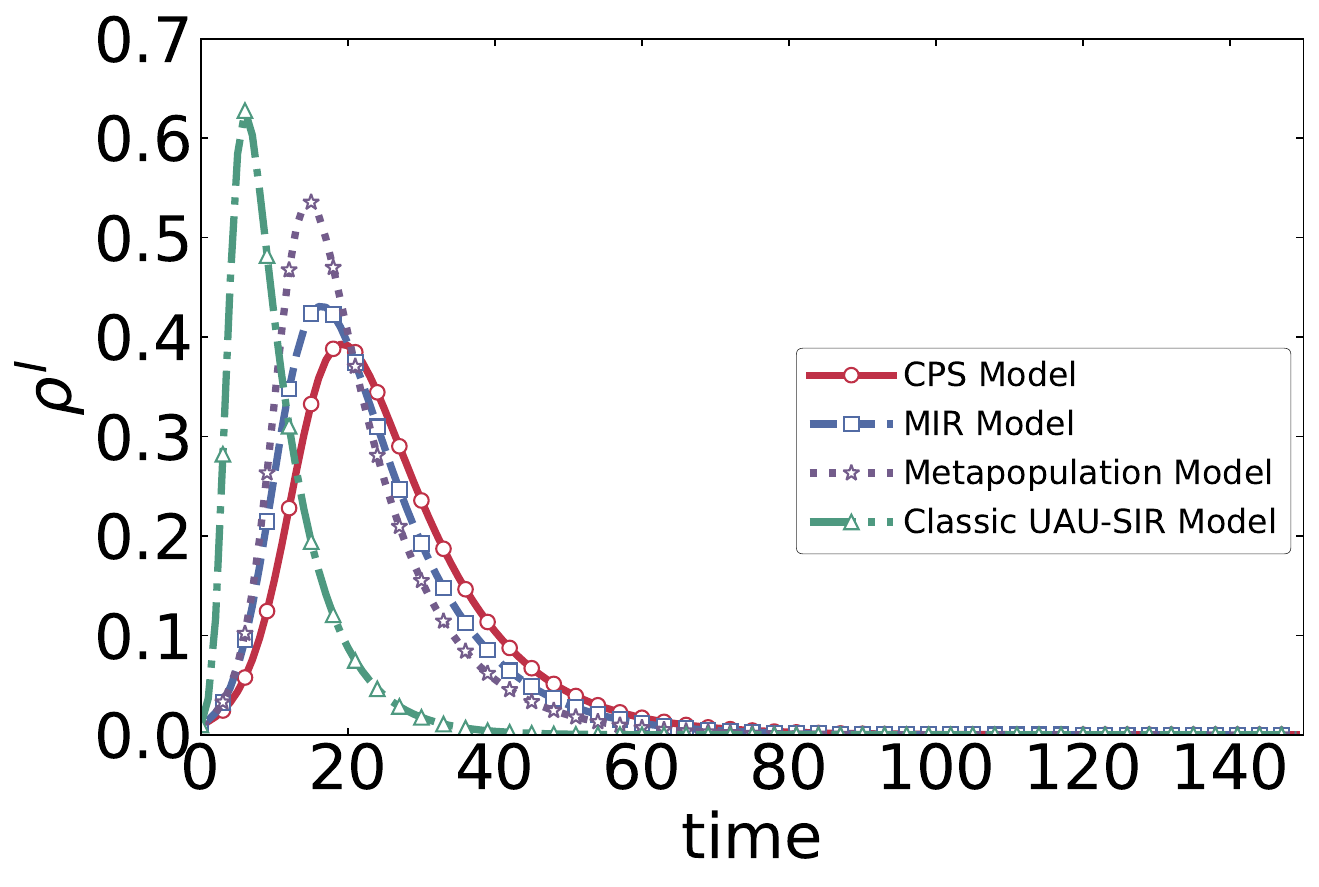}
    \caption{}
    \label{test8_2}
  \end{subfigure}
  \hspace*{-0.8em}
  \begin{subfigure}[b]{0.34\textwidth}
    \centering
    \includegraphics[width=\textwidth]{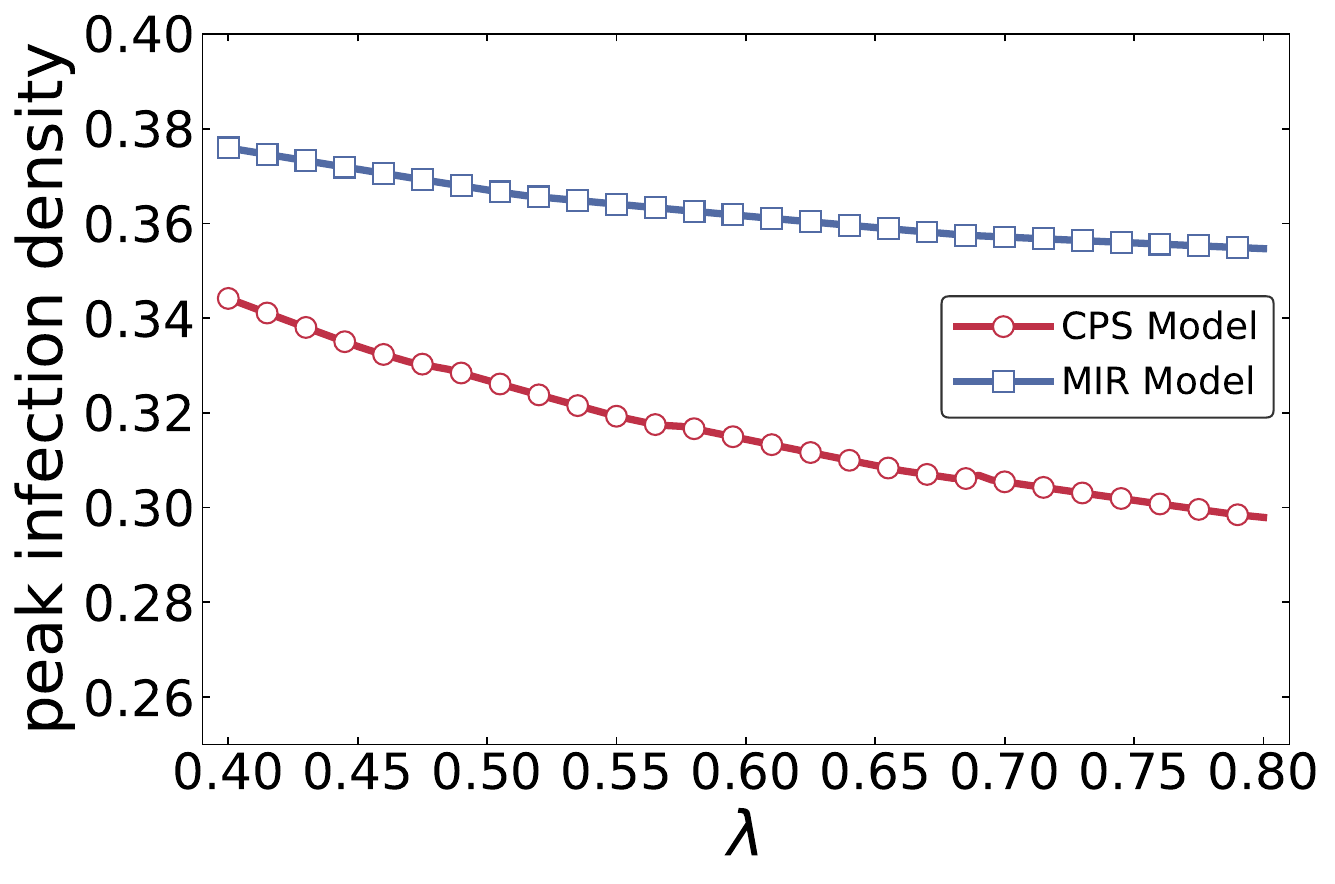}
    \caption{}
    \label{test8_3}
  \end{subfigure}
  \caption{
    \textbf{Comparative dynamics of awareness and epidemic spreading across the proposed CPS model and classical baselines.}
    (a) Temporal evolution of the global awareness density $\rho^A$.
    (b) Temporal evolution of the global infection density $\rho^I$.
    (c) Peak infection densities as a function of the information transmission rate $\lambda$.
    The classic UAU-SIR model employs a multiplex network with average degree $\langle k \rangle =10$ and total population $3000$. The metapopulation SIR, MIR, and CPS models each consist of $N=15$ patches with $200$ individuals per patch (plus $M=5$ transit hubs in CPS). Mobility probability is fixed at $g=0.2$ for MIR and CPS, with awareness-triggered migration parameters $\epsilon_0=0.5$ and $\alpha=0.5$ for CPS. Information and disease layers have comparable average degree ($\langle k \rangle =10$ where applicable).
  }
  \label{fig_8}
\end{figure*}

Finally, we assess the structural role of transfer stations as rendezvous nodes that mediate cross-patch interactions.
In Fig.~\ref{fig_7}, we examine the peak infection density as a function of (a) the disease spreading rate \(\beta\), and (b) the information transmission rate \(\lambda\), under different configurations of the number of residences \(N\) and transfer stations \(M\), ensuring a total of 30 patches.

In Fig.~\ref{fig_7}(a), infection peaks rise monotonically with increasing \(\beta\), but the magnitude varies significantly across configurations. Specifically, systems with fewer transfer stations (\(M = 10\)) exhibit substantially higher infection peaks, indicating that concentrated movement through limited hubs intensifies transmission. Conversely, a larger number of transfer stations (\(M = 20\)) helps spatially disperse interactions and limits outbreak severity.
Fig.~\ref{fig_7}(b) reinforces this conclusion under varying \(\lambda\), where the most effective suppression occurs in configurations with more transfer stations, while systems with \(M = 10\) show limited responsiveness to information diffusion. These findings validate the rendezvous effect hypothesis when hubs are scarce, they become bottlenecks of disease amplification. Increasing the granularity of rendezvous structures enhances both spatial resilience and responsiveness to awareness-driven migration.

\subsection{Benchmarking with Classical Metapopulation and Awareness Models}

To further highlight the distinct contribution of the proposed cyber--physical epidemic framework,
we conduct additional benchmark simulations by comparing the CPS model with three classical baseline models represented in Fig.~\ref{fig_8}:
1) a metapopulation SIR model without awareness (Metapopulation Model),
2) a classical multiplex UAU--SIR awareness--epidemic coupling model without migration (Classical UAU-SIR Model), and
3) a standard MIR metapopulation model with recurrent mobility (MIR Model).
All models share consistent population scales and network connectivities, which enable a mechanism-level evaluation of how awareness diffusion and adaptive mobility jointly reshape epidemic outcomes.

In Fig.~\ref{fig_8}(a),
the classical UAU--SIR framework exhibits the fastest growth and the highest peak,
as awareness spreads in a fully mixed and spatially unconstrained population.
The MIR model yields a lower peak due to recurrent mobility, which homogenizes
patch-level information and weakens local awareness aggregation.
In contrast, the CPS model shows the slowest increase and the lowest saturation level:
awareness-triggered migration is activated only after local awareness exceeds
the threshold $\alpha$, inducing non-recurrent redistribution that suppresses
excessive awareness concentration in highly informed patches.
Note that the metapopulation SIR model does not include an awareness layer at all.

Fig.~\ref{fig_8}(b) reports the corresponding infection dynamics.
The classical UAU--SIR model reaches the earliest and highest epidemic peak,
while the metapopulation model achieves only limited suppression via spatial fragmentation without behavioral feedback.
The MIR model further reduces the peak through awareness-induced protection, yet its mobility remains non-adaptive.
The CPS framework achieves the lowest peak and the longest outbreak duration,
reflecting the synergistic effect of behavioral awareness and adaptive,
event-triggered migration that actively redistributes individuals away from infection hotspots.

Fig.~\ref{fig_8}(c) shows the peak infection density as a function of the
information transmission rate $\lambda$.
Increasing $\lambda$ monotonically suppresses epidemic peaks in both CPS and MIR models.
However, the CPS model consistently outperforms the MIR baseline, with the performance
gap widening at higher $\lambda$.
This advantage stems from the fact that stronger information diffusion not only reduces susceptibility but also accelerates the activation of awareness-triggered migration, translating information into adaptive mobility responses that are absent
in recurrent movement models.

\section{Conclusion}
\label{sec5:conclusion}

This work proposed a cyber–physical epidemic framework that couples information diffusion with epidemic dynamics on a bipartite metapopulation network. By distinguishing residences from transfer stations, the model captures the rendezvous effect of transient gatherings. An awareness-triggered migration mechanism enables adaptive mobility when local awareness exceeds a threshold. Using microscopic Markov chains, we derive epidemic thresholds and validated them via Monte Carlo simulations. Comparative benchmarks against three classical models show that event-triggered migration yields stronger epidemic suppression than recurrent mobility or awareness-only mechanisms. Our results reveal how information transmission and mobility jointly shape outbreak risk and peak infection levels.
The framework extends classical metapopulation models by embedding feedback between awareness, adaptive movement, and spatial mixing. It offers a tractable tool for modeling behavioral interventions in cyber–physical public health systems, including risk communication and mobility-based mitigation strategies.

Nevertheless, several limitations remain.
First, the current mobility formulation mainly describes short-range movements and does not yet capture other realistic mobility patterns, such as long-range travel and multi-scale commuting.
Second, the present patch-based metapopulation resolution does not explicitly address finer spatial scales or continuous-space constraints.
Third, although the cyber-layer topology is assumed static for analytical tractability, real information diffusion is often governed by adaptive and time-varying communication patterns. Incorporating dynamic networks, along with higher-order group interactions such as simplicial complexes, is an important direction for improving the behavioral realism of the framework.

Future work will aim to incorporate richer mobility types, dynamic awareness structures, and multi-scale spatial modeling to better reflect real-world epidemic CPS scenarios.
Overall, this study provides a step toward analytically grounded digital epidemiology models where awareness, mobility adaptation, and infection co-evolve in networked systems.


%





\ifCLASSOPTIONcaptionsoff
  \newpage
\fi

\bibliographystyle{IEEEtran}
\bibliography{IEEEexample}

\begin{IEEEbiography}[{\includegraphics[width=1in,height=1.25in,clip,keepaspectratio]{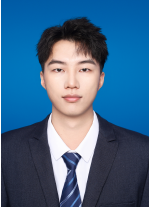}}]{Yusheng Li} received the B.E. degree from the School of Computing and Software, University of South China, Hunan, China. He is currently pursuing the M.S. degree at the College of Artificial Intelligence, Southwest University, Chongqing, China. 

His research interests include Complex Networks, Mathematical Epidemiology, Stochastic Processes, and Nonlinear Science.
\end{IEEEbiography}

\begin{IEEEbiography}[{\includegraphics[width=1in,height=1.25in,clip,keepaspectratio]{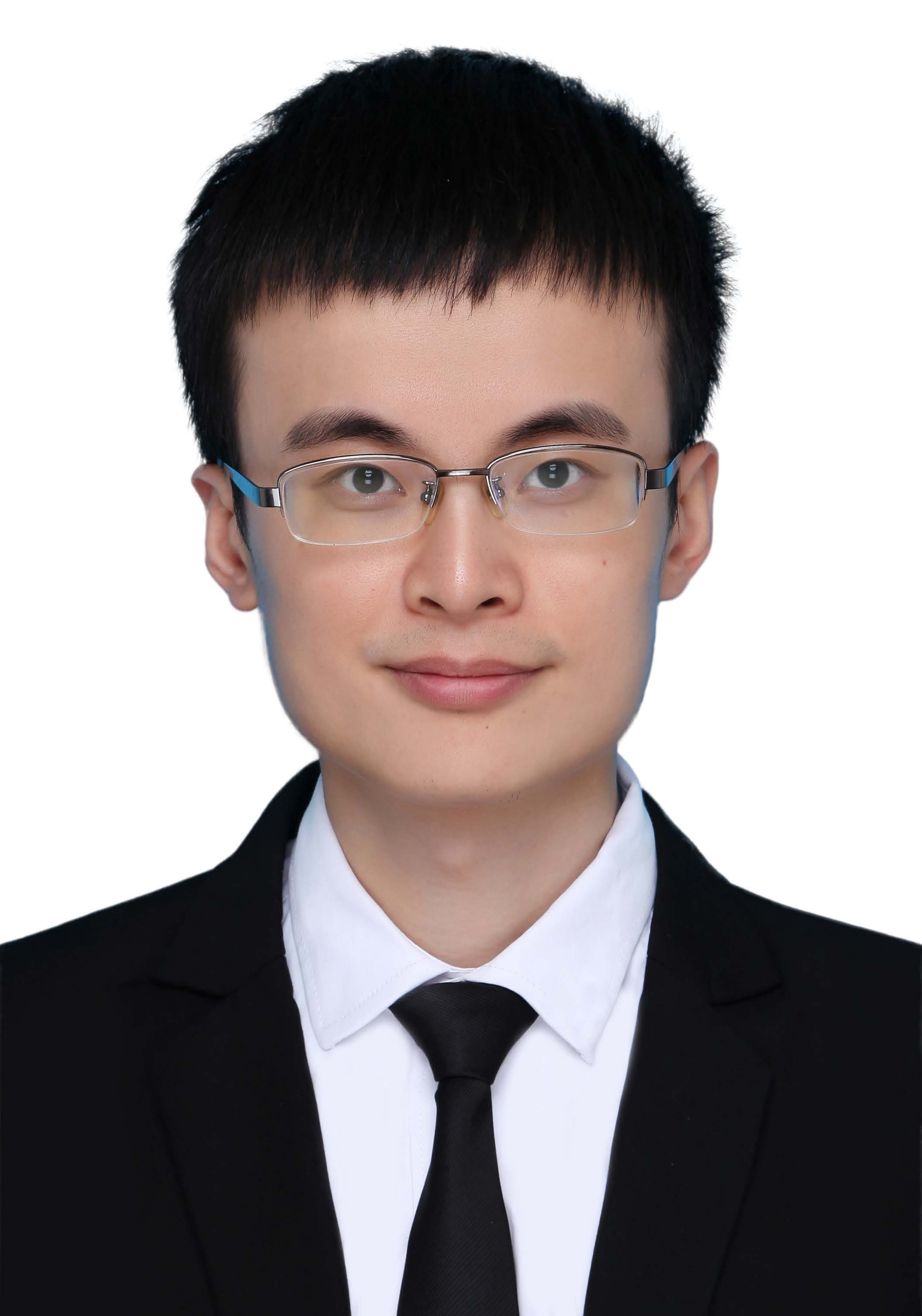}}]{Minyu Feng} (IEEE Senior Member) received his Ph.D. degree in Computer Science from a joint program between University of Electronic Science and Technology of China, Chengdu, China, and Humboldt University of Berlin, Berlin, Germany, in 2018. Since 2019, he has been an associate professor at the College of Artificial Intelligence, Southwest University, Chongqing, China. 
Dr. Feng has published more than 80 peer-reviewed papers in authoritative journals, such as IEEE Transactions on Pattern Analysis and Machine Intelligence, IEEE Transactions on Systems, Man, and Cybernetics: Systems, IEEE Transactions on Cybernetics, etc. He is a Senior Member of China Computer Federation (CCF) and Chinese Association of Automation (CAA). 

Currently, he serves as a Subject Editor for Applied Mathematical Modelling, an Academic Editor for PLOS Computational Biology, an Editorial Advisory Board Member for Chaos, and an Editorial Board Member for Humanities \& Social Sciences Communications, Scientific Reports, and International Journal of Mathematics for Industry. Besides, he is a Reviewer for Mathematical Reviews of the American Mathematical Society.

Dr. Feng's research interests include Complex Systems, Evolutionary Game Theory, Computational Social Science, and Mathematical Epidemiology.
\end{IEEEbiography}

\begin{IEEEbiography}[{\includegraphics[width=1in,height=1.25in,clip,keepaspectratio]{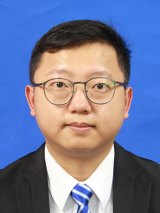}}]{Liang-jian Deng} (Senior Member, IEEE) received the B.S. and Ph.D.
degrees in applied mathematics from the School of Mathematical
Sciences, University of Electronic Science and Technology of China
(UESTC), Chengdu, China, in 2010 and 2016, respectively. He is currently a Research Fellow with the School of Mathematical Sciences,
UESTC. From 2013 to 2014, he was a Joint-Training Ph.D. student with
the Case Western Reserve University, Cleveland, OH, USA. In 2017, he
was a Postdoc with Hong Kong Baptist University (HKBU). In addition,
he also stayed at Isaac Newton Institute for Mathematical Sciences,
Cambridge University and HKBU for short visits. 

His research interests include Data Fusion, Image Processing, Deep Learning, Variational Modelling and Algorithms, and Numerical PDE.
\end{IEEEbiography}

\begin{IEEEbiography}[{\includegraphics[width=1in,height=1.25in,clip,keepaspectratio]{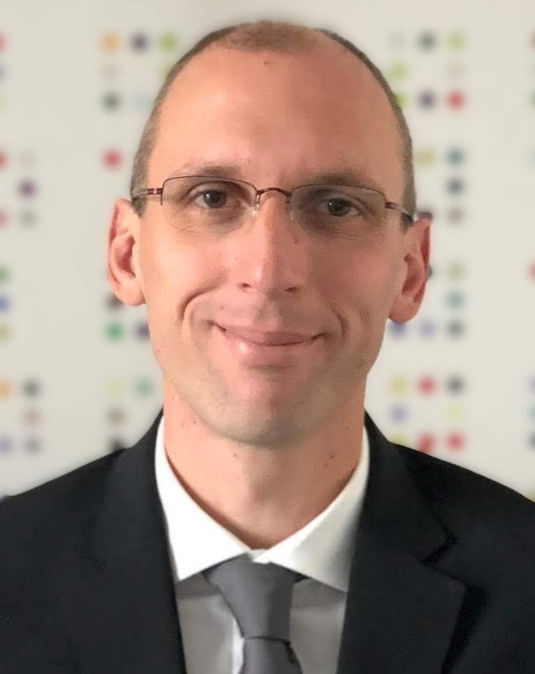}}]{Matjaž Perc} Matjaž Perc (Member, IEEE) received the Ph.D. degree in physics from the University of Maribor in 2006. He is currently Professor of Physics at the University of Maribor, Staff Researcher at the Community Healthcare Center Dr. Adolf Drolc Maribor, and Adjunct Professor at Kyung Hee University and Korea University. He is a member of Academia Europaea and the European Academy of Sciences and Arts, and among top 1\% most cited physicists according to Clarivate Analytics data. He is also the 2015 recipient of the Young Scientist Award for Socio and Econophysics from the German Physical Society, and the 2017 USERN Laureate. In 2018 he received the Zois Award, which is the highest national research award in Slovenia. In 2019 he became Fellow of the American Physical Society.
\end{IEEEbiography}

\begin{IEEEbiography}[{\includegraphics[width=1in,height=1.25in,clip,]{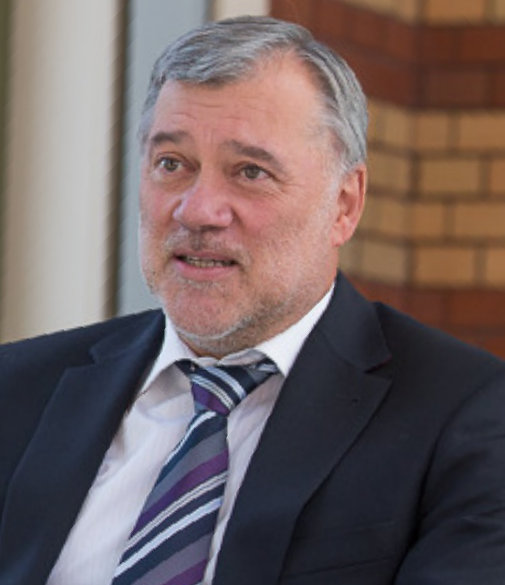}}]{J{\"u}rgen Kurths} received the B.S. degree in mathematics from the University of Rostock, Rostock, Germany, the Ph.D. degree from the Academy of Sciences, German Democratic Republic, Berlin, Germany, in 1983, the Honorary degree from N.I.Lobachevsky State University, Nizhny Novgorod, Russia, in 2008, and the Honorary degree from Saratow State University, Saratov, Russia, in 2012. From 1994 to 2008, he was a Full Professor with the University of Potsdam, Potsdam, Germany. Since 2008, he has been a Professor of Nonlinear Dynamics with the Humboldt University of Berlin, Berlin, Germany, and the Chair of the Research Domain Complexity Science with the Potsdam Institute for Climate Impact Research, Potsdam. 

He has authored more than 700 papers, which are cited more than 60,000 times (H-index: 111). His main research interests include Synchronization, Complex Networks, Time Series Analysis, and their Applications. Dr. Kurths was the recipient of the Alexander von Humboldt Research Award from India, in 2005, and from Poland in 2021, the Richardson Medal of the European Geophysical Union in 2013, and the Eight Honorary Doctorates. He is a Highly Cited Researcher in Engineering. He is a member of the Academia Europaea. He was an Editor-in-Chief of Chaos and currently serves on the editorial boards of more than ten journals. He is a Fellow of the American Physical Society, the Royal Society of Edinburgh, and the Network Science Society.
\end{IEEEbiography}



\end{document}